\documentclass[useAMS,usenatbib]{mn2e}
\usepackage{amsmath}
\usepackage{txfonts}
\usepackage{graphics}
\usepackage{times}

\newcommand{\mnras}{MNRAS}

\newcommand{\apjl}{ApJL}

\newcommand{\nat}{Nature}

\begin{document}

% --- title --- %
\title[Redshift space distortions]{Towards an accurate model of the redshift
space clustering of halos in the quasilinear regime}

\author[Reid \& White]{
Beth A. Reid$^{1,2}$\thanks{E-mail: beth.ann.reid@gmail.com}, Martin White$^{1,3}$ \\
$^{1}$ Lawrence Berkeley National Laboratory, 1 Cyclotron Road, Berkeley, CA 94720, USA \\
$^{2}$ Hubble Fellow \\
$^{3}$ Departments of Physics and Astronomy, University of California, Berkeley, CA 94720, USA
}

\date{\today} 
\pagerange{\pageref{firstpage}--\pageref{lastpage}}

\maketitle

\label{firstpage}

\begin{abstract}
Observations of redshift-space distortions in spectroscopic galaxy surveys offer an attractive method for measuring the build-up of cosmological structure, which depends both on the expansion rate of the Universe and our theory of gravity.  The statistical precision with which redshift space distortions can now be measured demands better control of our theoretical systematic errors.  While many recent studies focus on understanding dark matter clustering in redshift space, galaxies occupy special places in the universe: dark matter halos.  In our  detailed study of halo clustering and velocity statistics in $67.5\,h^{-3}{\rm Gpc}^3$ of $N$-body simulations, we uncover a complex dependence of redshift space clustering on halo bias.  We identify two distinct corrections which affect the halo redshift space correlation function on quasilinear scales ($\sim 30-80\,h^{-1}$Mpc): the non-linear mapping between real and redshift space positions, and the non-linear suppression of power in the velocity divergence field.  We model the first non-perturbatively using the scale-dependent Gaussian streaming model, which we show is accurate at the $<0.5$ (2) per cent level in transforming real space clustering and velocity statistics into redshift space on scales $s>10$ ($s>25$)$h^{-1}$Mpc for the monopole (quadrupole) halo correlation functions.  The dominant correction to the Kaiser limit in this model scales like $b^3$.  We use standard perturbation theory to predict the real space pairwise halo velocity statistics.  Our fully analytic model is accurate at the 2 per cent level only on scales $s > 40\,h^{-1}$Mpc for the range of halo masses we studied (with $b=1.4-2.8$).  We find that recent models of halo redshift space clustering that neglect the corrections from the bispectrum and higher order terms from the non-linear real-to-redshift space mapping will not have the accuracy required for current and future observational analyses.  Finally, we note that our simulation results confirm the essential but non-trivial assumption that on large scales, the bias inferred from real space clustering of halos is the same one that determines their pairwise infall velocity amplitude at the per cent level.
\end{abstract}


\section{Introduction}

The growth of large-scale structure, as revealed in the clustering of
galaxies observed in large redshift surveys, has historically been one
of our most important cosmological probes.  This growth is driven by a
competition between gravitational attraction and the expansion of space-time,
allowing us to test our model of gravity and the expansion history of the
Universe.  Despite the fact that galaxy light doesn't faithfully trace the
mass, even on large scales, galaxies are expected to act nearly as test
particles within the cosmological matter flow.  Thus the motions of galaxies
carry an imprint of the rate of growth of large-scale structure and allows
us to both probe dark energy and test General Relativity
\cite[see e.g.][for recent studies]{jain08,NesPer08,Song08a,Song08b,PerWhi08,McDSel08,WhiSonPer09}.

This measurement of the growth of structure relies on redshift-space
distortions seen in galaxy surveys \citep{Kai87}.  Even though we expect the
clustering of galaxies in real space to have no preferred direction, galaxy
maps produced by estimating distances from redshifts obtained in spectroscopic
surveys reveal an anisotropic galaxy distribution.
This anisotropy arises because galaxy recession velocities, from which
distances are inferred, include components from both the Hubble flow and
peculiar velocities driven by the clustering of matter
\citep[see][for a review]{HamiltonReview}.
Measurements of the anisotropies allow constraints to be placed on the
rate of growth of clustering.

On large scales, where linear perturbation theory is valid, it is natural
to work in a Fourier basis because the symmetries of the background solution imply
that $\mathbf{k}$-modes evolve independently.  
On smaller scales, and especially once survey non-idealities
and fingers-of-god become important, the choice is not so clear.  
Because the velocity field departs from its linear theory prediction on extremely large scales
($k \lesssim 0.03\,h\,{\rm Mpc}^{-1}$), models beyond linear theory must be used to extract
cosmological information from redshift surveys.
This has been long recognised and a variety of methods have been attempted to model the
distortions.  Several recent studies of redshift space distortions have provided non-linear
descriptions of the matter density field in Fourier space \citep{TarNisSai10,jennings/etal:2011a}
which agree well with direct N-body calculations of the effect.
However, we do not generally observe the matter density but rather tracers which tend to live in dark matter halos.  This introduces further effects which must be carefully modelled if we are to achieve the desired accuracy.
In this paper we find a strong dependence on halo bias in the shape of the redshift space correlation function, indicating the need for more sophisticated theoretical models \citep[see also ][for a Fourier-space approach]{TarSaiNis11}.  We trace this strong bias dependence primarily to the non-linear mapping between real and redshift space.  While more slowly varying with bias, non-linear evolution of the pairwise velocity distribution also substantially changes the redshift space clustering in comparison to linear theory.

\citet{tinker/weinberg/zheng:2006} and \citet{tinker:2007}, building on the work of \citet{HatCol99}, combined the streaming and halo models to describe the redshift space correlation function on scales of $r<20\,h^{-1}$Mpc.  While we take a similar approach here, we focus on larger scales and only on halos and ignore the contributions from satellite galaxies for now.  This greatly simplifies our modelling compared with \citet{tinker:2007}; however, as we will see, there is still a rich phenomenology compared to dark matter clustering.

The outline of the paper is as follows.  We first review linear and quasilinear descriptions of redshift space distortions in both Fourier and configuration space, and introduce the scale-dependent Gaussian streaming model that we study in detail in this paper.  In Section \ref{fisher} we present a simplified Fisher matrix calculation of the expected constraint on the peculiar velocity field from the ongoing Baryon Oscillation Spectroscopic Survey \citep[BOSS;][]{schlegel/white/eisenstein:2009,eisenstein/etal:2011}.  This calculation sets the target for the accuracy of our model.  In Section \ref{simxis} we compare the streaming model ansatz as a transformation between real space clustering and velocity statistics and clustering in redshift space for halo mass bins, finding very good agreement.  Then Section \ref{ptsec} examines the ability of perturbation theory to describe the four ingredients of the streaming model ansatz for tracers of the matter density field: the real space correlation function, the mean tracer pairwise velocity, and the tracer velocity dispersions along and perpendicular to the line-of-sight (LOS).  We show in Section \ref{combinesec} that using our perturbation theory description as input into the scale-dependent Gaussian streaming model, we have an analytic model accurate at the 2 per cent level on scales $s > 40\,h^{-1}$Mpc.  Sections \ref{zdep} and \ref{biasdep} identify the dominant non-linear terms, elucidating how our model depends on redshift and halo bias.

\section{Modelling Redshift Space Distortions}

We begin by reviewing the effect of redshift space distortions in linear theory both in Fourier space and  configuration space.  While Fourier space is usually preferred for theoretical investigations, configuration space is simpler when dealing with wide-angle effects \citep[e.g.][]{PapSza08} and often preferred in observational analyses.

\subsection{Linear theory: Fourier space}
\label{background}

The redshift-space position {\bf s} of a galaxy differs from its real-space
position {\bf r} due to its peculiar velocity,
\begin{equation} \label{eq:sx}
  {\bf s} = {\bf x} + v_z({\bf x})\,\widehat{\bf z},
\end{equation}
where $v_z({\bf x}) \equiv u_z({\bf x})/(aH)$ is the line-of-sight (LOS) component of the galaxy
velocity (assumed non-relativistic) in units of the Hubble velocity,
and we have taken the LOS to be the $z$-axis.
We shall adopt the ``plane-parallel'' approximation, so this
direction is fixed for all tracers (halos, galaxies, etc.).

The galaxy over-density field in redshift-space can be obtained by
imposing mass conservation, $(1+\delta_g^s)d^3s=(1+\delta_g)d^3r$.  For a uniform, $z$-independent mean galaxy density,  
the exact Jacobian for the real-space to redshift-space transformation is
\begin{equation}
  \frac{d^3s}{d^3r} = \left(1+\frac{v_z}{z}\right)^2
    \left(1+\frac{dv_z}{dz}\right).
\end{equation}
In the limit where we are looking at scales much smaller than the mean distance
to the pair, $v_z/z$ is small and it is only the second term that is important
(\citealt{Kai87}; but see \citealt{PapSza08,ShaLew08}),
\begin{equation}
  1+\delta_g^s = \left(1+\delta_g\right)\left(1+\frac{dv_z}{dz}\right)^{-1}.
\end{equation}
If we assume an irrotational velocity field we can write
$v_z= -\partial/\partial z\,\nabla^{-2}\theta$,
where $\theta\equiv-\nabla\cdot{\bf v}$, and $\nabla^{-2}$ is the inverse
Laplacian operator. In Fourier space,
$(\partial/\partial z)^2\nabla^{-2}=(k_z/k)^2=\mu^2$, where $\mu$ is the
cosine of the LOS angle, so we have that
\begin{equation}
  \delta_g^s(k) = \delta_g(k) + \mu^2\theta(k),
\end{equation}
to linear order.
Often it is further assumed that the velocity field comes from linear
perturbation theory. Then
\begin{equation}
 \theta(k) = f \delta_{\rm mass}(k),
\label{eq:thetadeltalin}
\end{equation}
where $f\equiv d\ln D/d\ln a \approx \Omega_m^{0.6}$ \citep{Pee80}.

For a population of galaxies, which we denote with a subscript $g$,
the linear, redshift-space power spectrum is then proportional to the
linear, real-space matter power spectrum, $P_m^r(k)$
\begin{equation}
 P_g^s(k,\mu) = \left(b+f\mu^2\right)^2 P_m^r(k)
               = b^2\left(1+\beta\mu^2\right)^2 P_m^r(k).
\label{pkkaiser}
\end{equation}

\subsection{Linear theory: configuration space}

\begin{figure}
\begin{center}
\resizebox{3.2in}{!}{\includegraphics{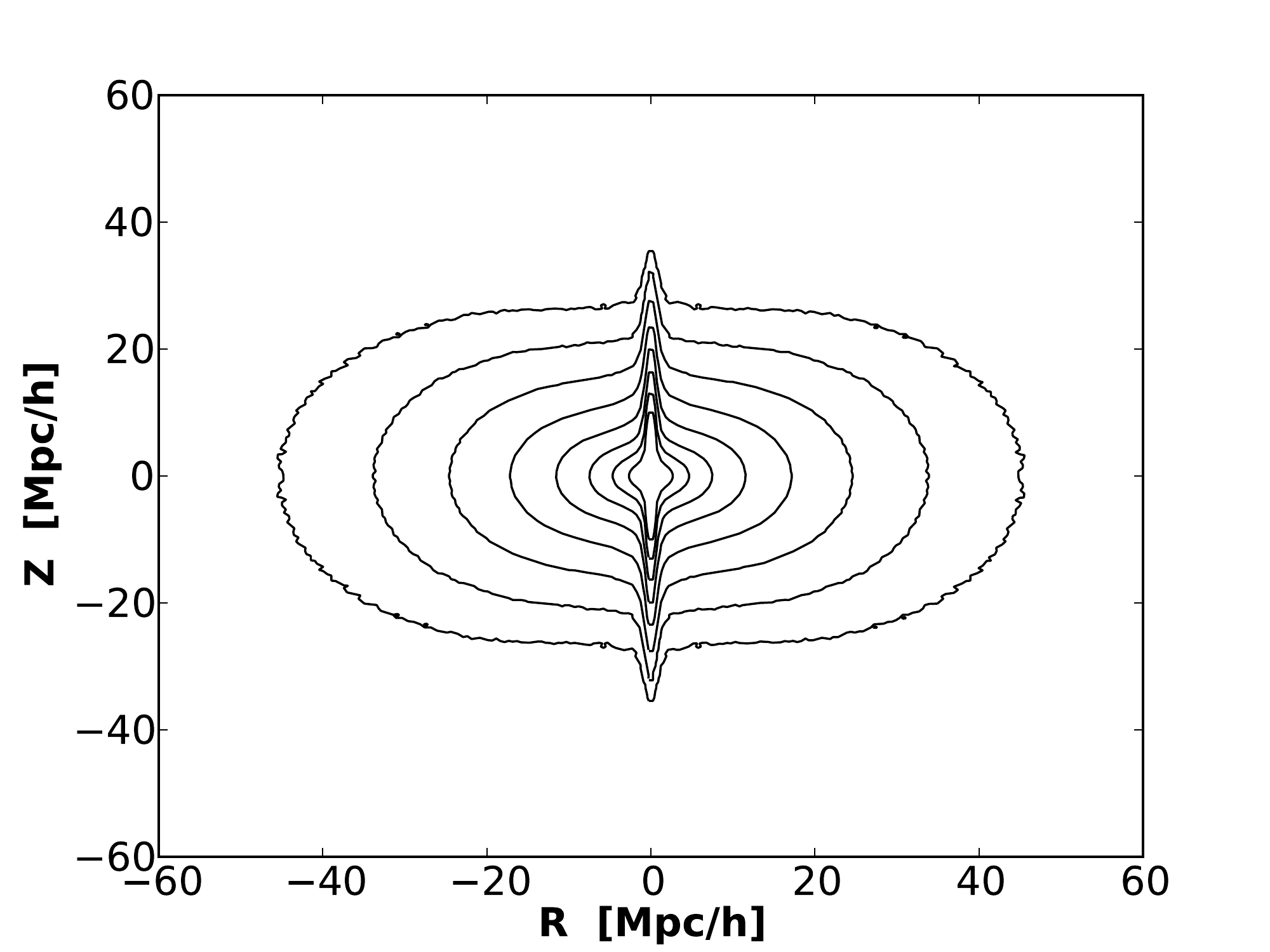}}
\end{center}
\caption{Contours of constant $\xi$ as a function of separation perpendicular (R) and parallel (Z) to the line-of-sight (LOS) from the mock galaxy catalogues in \citet{white/etal:2011}.  The magnitude of the squashing along the Z axis depends on the amplitude of the peculiar velocity field, which is related to the over-density field on large scales (Eq.~\ref{eq:thetadeltalin}).  The stretching of the contours near R=0 results from small-scale velocity dispersions and is often referred to as the `fingers-of-god.'}
\label{fig:butterfly}
\end{figure}

\citet{Fis95} first showed in detail the relation between the Kaiser formula in Fourier space, Eq.~\ref{pkkaiser}, and the redshift space correlation function; we rely heavily on that work for this section.
In linear theory, the correlation between $\delta({\bf x})$ and ${\bf v}({\bf x'})$ gives rise to a mean infall $v_{12}({\bf r})$ between pairs of matter tracers.  The velocity dispersion along the LOS, $\sigma_{12}^2(r) = \left\langle ({\bf v}_{z}({\bf x}) - {\bf v}_{z}({\bf x'})^2\right\rangle$, depends both on scale and the orientation of the pair separation vector with respect to the LOS.  These scale dependencies give rise to linear redshift-space distortions in configuration space, apparent in Fig.~\ref{fig:butterfly} as the squashing of contours along the $Z$ axis.

We reproduce the relevant linear theory velocity predictions here, but generalised to linearly biased tracers with bias $b$:
\begin{eqnarray}
{\bf v}_{12}(r) = v_{12}(r) \hat{r} = -\hat{r} \frac{fb}{\pi^2} \int dk\ k\ P_m^r(k) j_1(kr) \label{linearinfall} \\
\left < {\bf v}_i({\bf r'}+\bf{r}) {\bf v}_j({\bf r'}) \right\rangle = \Psi_{\perp}({\bf r}) \delta^{K}_{ij} + [\Psi_{\parallel}(r) - \Psi_{\perp}(r)] \hat{r}_i \hat{r}_j \label{defvcorrlin} \\
\Psi_{\perp}(r) = \frac{f^2}{2\pi^2} \int dk\ P_m^r(k) \frac{j_1(kr)}{kr}\label{psiperp} \\
\Psi_{\parallel}(r) = \frac{f^2}{2\pi^2} \int dk\ P_m^r(k) \left[j_0(kr) - \frac{2j_1(kr)}{kr} \right] \label{psiparallel} \\
\sigma_{12}^2(r, \mu^2) = 2\left[ \sigma_v^2 - \mu^2 \Psi_{\parallel}(r) - (1-\mu^2) \Psi_{\perp}(r)\right] \label{vdisptot}
\end{eqnarray}
where $\mu = \hat{\ell} \cdot \hat{r}$ denotes the angle between the LOS and the pair separation vector and $\sigma_v^2$ is the one-dimensional velocity dispersion, $\frac{1}{3}\left\langle{\bf v}({\bf x}) \cdot {\bf v}({\bf x})\right\rangle$.  Our Fourier convention is
\begin{equation} \label{eq:fourier}
f({\bf x}) = \int \frac{d^3k}{(2\pi)^3} e^{i {\bf k} \cdot {\bf x}} \tilde{f}({\bf k}); \; \tilde{f}({\bf k}) = \int d^3xe^{-i {\bf k} \cdot {\bf x}} \tilde{f}({\bf k}).
\end{equation}

\citet{Fis95} derived the linear theory redshift space distortion limit in configuration space under the assumption that the density and velocity fields are Gaussian, which amounts to evaluating the following expression:
\begin{equation}
    1 + \xi^s_g(r_\sigma,r_\pi)
        = \left\langle \int dy\  \left(1 + \delta_1\right)\left(1 + \delta_2\right)
            \delta^D(r_\pi - y + v_1 - v_2) \right\rangle,
\label{eqn:configremap}
\end{equation}
where $\delta_{1/2}$ and $v_{1/2}$ are the density and LOS velocity at two points with real space separation $y$ along the LOS and $r_{\sigma}$ perpendicular to the LOS, and $\delta^D$ is the Dirac delta function ensuring that the pair is mapped to redshift space separation $r_{\pi}$.  We can then re-express $\delta^D$
\begin{equation}
  1 + \xi^s_g(r_\sigma,r_\pi) = \left\langle \int dy \left(1 + \delta_1\right)\left(1 + \delta_2\right)
  \int \frac{d\kappa}{2\pi}\ e^{i\kappa(r_\pi - y + v_1 - v_2)} \right\rangle
\label{eqn:allorders}
\end{equation}
and compute the expectation value, assuming Gaussian statistics \citep[eq.~20 of ][]{Fis95}:
\begin{eqnarray}
   1 + \xi^s_g(r_\sigma,r_\pi) = \int \frac{dy}{\sqrt{2\pi\sigma_{12}^2(y)}} \exp\left[-\frac{(r_\pi-y)^2}{2\sigma_{12}^2(y)}\right] \times \nonumber \\
    \left[ 1 + \xi^r_g(r) + \frac{y}{r} \frac{(r_\pi-y)v_{12}(r)}{\sigma_{12}^2(y)} - \frac{1}{4} \frac{y^2}{r^2} \frac{v_{12}^2(r)}{\sigma_{12}^2(y)} \left(1 - \frac{(r_\pi-y)^2}{\sigma_{12}^2(y)}\right) \right]. \label{eq:exactgaussian}
\end{eqnarray}
If we expand Eq.~\ref{eq:exactgaussian} to linear order, the redshift space correlation function is equivalent to 
\begin{equation}
  \xi^s_g(r_{\sigma},r_{\pi}) = \xi_g^r(s) - \left.\frac{d}{dy} \left[ v_{12}(r) \frac{y}{r} \right] \right|_{y=r_{\pi}} +   \frac{1}{2}\left. \frac{d^2}{dy^2} \left[\sigma_{12}^2(y)\right] \right|_{y=r_{\pi}},
\label{fisherlinearxi}
\end{equation}
where $\xi_g^r(s) = b^2 \xi_m^r(s)$ is the real space linear galaxy correlation function evaluated at redshift space separation $s^2=r_\sigma^2 + r_\pi^2$ and $r^2 = r_{\sigma}^2 + y^2$ is the real space separation of the pair.  We expand the last two terms to elucidate the contribution from different halo velocity statistics as a function of angle and separation.  The terms proportional to $bf$ in Eq. \ref{pkkaiser} arise from the pairwise infall and its derivative as a function of separation $r$ (denoted with $'$ throughout), and terms proportional to $f^2$ arise from the pairwise velocity dispersion and its first and second derivatives:
\begin{eqnarray}
-\frac{d}{dy} \left.\left[ v_{12}(r) \frac{y}{r} \right] \right|_{y=r_{\pi}} = \frac{v_{12}(r)}{r} (\mu^2-1) - v_{12}'(r) \mu^2 \label{vexpand} \\
 \frac{1}{2} \frac{d^2}{dy^2} \left.\left[\sigma_{12}^2(y,r_{\sigma})\right]\right|_{y=r_{\pi}} = (2-10\mu^2 + 8\mu^4) \frac{\Psi_{\perp}(r) - \Psi_{\parallel}(r)}{r^2} \nonumber\\
+(-5\mu^2 + 5\mu^4) \frac{\Psi'_{\parallel}(r)}{r} + (-1+6\mu^2 - 5\mu^4) \frac{\Psi'_{\perp}(r)}{r} \nonumber\\
-\mu^4 \Psi''_{\parallel}(r) + (-\mu^2 + \mu^4) \Psi''_{\perp}(r). \label{vdispexpand}
\end{eqnarray}
Since the theoretical prediction for the redshift space correlation function depends on first and second derivatives of velocity statistics, scale-dependent errors on a theoretical model for these functions can translate into large errors on $\xi^{s}$.  Eq.~\ref{vdispexpand} also demonstrates that at linear order, any constant, isotropic velocity dispersion does not alter the redshift space correlation function, since only the difference $\Psi_{\perp}(r) - \Psi_{\parallel}(r)$ enters, along with derivatives of those functions.

\subsection{Moments}

It is standard practise to expand the dependence of $P(\mathbf{k})$ or
$\xi(\mathbf{s})$ on the LOS angle in Legendre polynomials,
which we here write as $L_\ell$ to avoid confusion with the power spectrum
moments:
\begin{equation}
  P(k,\mu_k) = \sum_\ell P_\ell(k) L_\ell(\mu_k).
\end{equation}
The power spectrum moments are given by
\begin{equation}
  P_\ell(k)= \frac{2\ell+1}{2}\int d\mu\ P(k,\mu_k)L_\ell(\mu_k).
\end{equation}
In linear theory, only $\ell =$ 0, 2, and 4 are non-zero.  The power spectrum is particularly simple because the $k$-dependencies of the Legendre moments $P_\ell(k)$ are identical, and the relative amplitudes depend only on the sample bias $b$ and the rate of structure growth $f$.
\begin{equation} \label{pmoments}
\begin{pmatrix}
P_0(k)\\
P_2(k)\\
P_4(k)
\end{pmatrix}
= P_m^r(k)
\begin{pmatrix}
b^2+\frac{2}{3} bf +\frac{1}{5} f^2\\
\frac{4}{3} bf + \frac{4}{7} f^2\\
\frac{8}{35} f^2
\end{pmatrix}
\end{equation}
Fourier transforming to configuration space gives \citep[e.g.][]{ColFisWei94,HamiltonReview}
\begin{equation} \label{eq:ximoments}
  \xi(s,\mu_s) = \sum_\ell \xi_\ell(s)L_\ell(\mu_s)
  \quad , \quad
  \xi_\ell(s) = i^\ell \int \frac{dk}{k} \Delta_\ell^2(k) j_\ell(ks)
\end{equation}
where $j_\ell$ is the spherical Bessel function of order $\ell$ and
$\Delta_\ell^2\equiv k^3P_\ell(k)/(2\pi^2)$ is the moment of the
dimensionless power spectrum or power per $\ln k$.  While the relative amplitudes of the Legendre moments $\xi_{\ell}(s)$ are also given by Eq.~\ref{pmoments}, in configuration space the moments depend differently on $s$ (see solid lines in Fig.~\ref{fig:streaminglin}).  For example, using the recurrence relations among the spherical Bessel functions we can write $\xi_2$ as the average value of $\xi$ up to $s$ minus the value at $s$.

Non-linear effects, and especially fingers-of-god on small scales,
can cause the $\ell > 4$ moments to be non-zero.  The fingers-of-god are evident in Fig.~\ref{fig:butterfly} as the stretching of contours localised near $R=0$.  The presence of
strong fingers-of-god also drives $\xi_2$ positive, whereas super-cluster
infall on larger scales generally predicts $\xi_2<0$.
Generally the monopole ($\ell=0$) and quadrupole ($\ell=2$) dominate
in terms of signal-to-noise, and the ratio $P_2/P_0$ (or an analogous combination of $\xi_0$ and $\xi_2$) encodes information
about $\beta=f/b$
\citep[but see][for a discussion of issues with this approach]{BerNarWei01}.

\subsection{Beyond linear theory}

\begin{figure}
\begin{center}
\resizebox{3.6in}{!}{\includegraphics{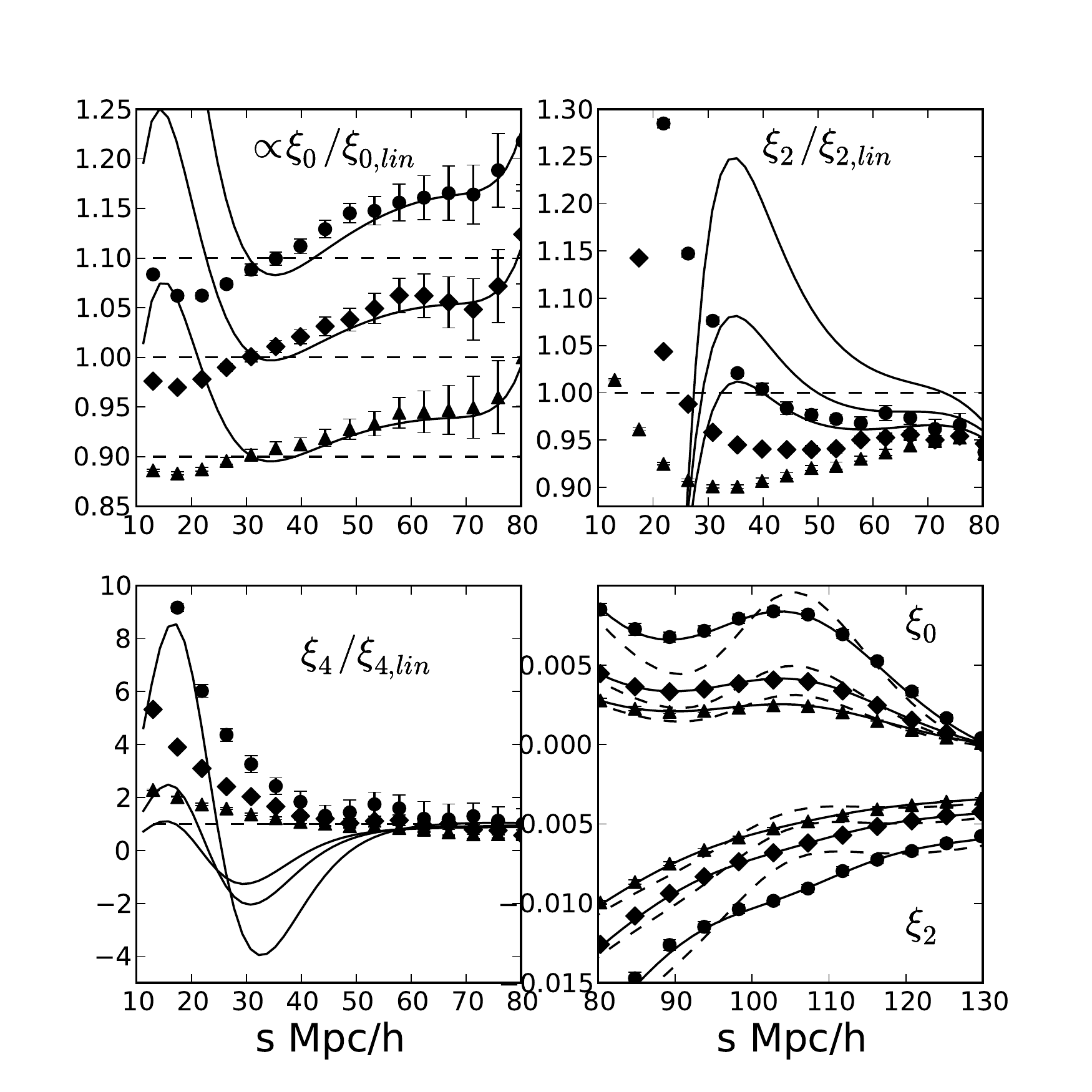}}
\end{center}
\caption{Comparison of $N$-body simulation redshift space clustering of halo samples in Table \ref{table:halos} with the biased Lagrangian perturbation theory of \citet{Mat08b}.  The high (circles) and low (triangles) halo mass bins are shifted by 10 per cent for clarity compared with the HOD subsample (diamonds) in the upper left panel.  The predictions of \citet{Mat08b} are shown as solid black curves; higher bias values have larger deviations from linear theory expectations (dashed curves) in LPT.  While \citet{Mat08b} provides an excellent fit to the simulation results for $\xi_0$ and $\xi_2$ on BAO scales (bottom right panel), the quasilinear scale description is not accurate.}
\label{fig:comparematsu}
\end{figure}

As is so often the case, our strongest statistical constraints come from
the smaller scales where we have a large number of independent samples
within our survey.  These are also the scales for which the simple linear
theory described in the last section is the least applicable.  
A standard approach to include small-scale non-linearities manifest as fingers-of-god 
is to use a streaming model, where linear
theory is spliced together with an approximation for random motion
of particles in collapsed objects \citep[e.g.][]{Pea92}.
This arises from ignoring the scale-dependence of the mapping between
real and redshift space separations and assuming an isotropic velocity
dispersion and amounts to convolving the linear theory result with a
LOS smearing or multiplying the power spectrum by the Fourier
transform of the small-scale velocity PDF.
\citet{Pea92} modelled the small-scale velocity field as an incoherent
Gaussian scatter which amounts to
\begin{equation} \label{eq:pkgauss}
  P_g^s(k,\mu) = \left(b+f\mu^2\right)^2 P_m^r(k)\, e^{-k^2\sigma^2\mu^2}.
\end{equation}
A more realistic distribution for the pairwise velocities is exponential
\citep{Pee76,DavPee83}, because galaxies populate halos of a wide range of
masses and velocity dispersions \citep{She96,DiaGel96,HaloRed,SeljakRed}.

While this extension from linear theory can improve agreement with observations and $N$-body simulation results, this simple approach ignores the well-known fact that the velocity divergence field $\theta$ deviates from linear theory on extremely large scales ($k \lesssim 0.03\,h\,{\rm Mpc}^{-1}$); see e.g., fig.~8 of \citet{carlson/white/padmanabhan:2009} for a recent comparison.
\citet{Sco04} therefore proposed a simple extension of linear theory for the matter power spectrum in redshift space:
\begin{equation} \label{scoccPT}
P_m^s(k,\mu) =\left[P^{r,PT}_{\delta\delta}(k) + 2f\mu^2 P^{r,PT}_{\delta \theta} + f^2\mu^4 P_{\theta \theta}^{r,PT}\right]e^{-k^2 \mu^2 \sigma_v^2}
\end{equation}
where the power spectra on the right hand side are evaluated in perturbation theory in {\em real} space, and $\sigma_v^2$ is the linear velocity dispersion.  \citet{jennings/etal:2011a,jennings/etal:2011b} have extended this model for the redshift space matter power spectrum to different dark energy scenarios by fitting a relation between the three non-linear power spectra in Eq.~\ref{scoccPT} to $N$-body simulation outputs, but treat $\sigma_v^2$ as a free parameter \citep[in a similar fashion to earlier work by][]{HatCol99}.

However, the relation between the real and redshift space statistics depends on
non-trivial correlations at high orders, which Eq.~\ref{scoccPT} ignores; many terms from the standard perturbation theory expression for the redshift space power spectrum have been dropped \citep[see Section IIIB and Appendix B of][]{Mat08a}.  However, it is well-known that the standard perturbation theory expression for the redshift space clustering of matter is inaccurate (e.g., \cite{SCF99}; see also \cite{Mat08a} and \cite{blake/etal:2011} for recent comparisons to $N$-body simulations and observed galaxy power spectra, respectively).
The importance of higher-order terms can be seen most easily from the expression for the correlation function, Eq.~\ref{eqn:allorders}, where we see that going beyond linear theory involves many moments of
$\delta$ and $v_z$.
\citet{TarNisSai10} showed that the bispectrum terms have oscillatory features which affect the BAO feature, and used them to obtain a better fit to $N$-body simulation data for the matter redshift space power spectrum.  
However, \citet{TarNisSai10} still needed to introduce a Gaussian damping term with free parameter $\tilde{\sigma}_v^2$ in order to fit the $N$-body data well.  \citet{tang/kayo/takada:2011} take a different approach, and try to reconstruct $P_{\delta \delta}^{r}(k)$ and $P_{\delta \theta}^{r}(k)$ from $P^s(k)$ for both matter and halos in $N$-body simulations, assuming a relation like Eq.~\ref{scoccPT}, but with a more general multiplicative damping term.  Their procedure does not work for halos unless they account for a higher order term in perturbation theory which scales like $b^2 B_{\delta \delta \theta}$.  

Going beyond linear theory perturbatively introduces extra ``mode coupling''
terms, some of which can be resummed into a $\mu$-dependent damping
\citep{Bha96,ESW,RPT,Mat08a,Mat08b,TarNisSai10}.
For example, in the formalism of
\citet{Mat08a,Mat08b} the damping is exponential with an angular dependence
$(\mathbf{k}+fk\mu\hat{\mathbf{z}})^2$ or
$[1+f(f+2)\mu^2]\,k^2$ and the mode-coupling terms go up to $\mu^8$.
However the range of validity of perturbation theory is limited, and it clearly cannot be extended to the dynamics of collapsed objects, so it is not clear how much guidance these forms provide.

Another issue of utmost importance in the application to galaxy surveys, but that was not addressed in most of these recent perturbation theory studies, is the inclusion of bias.  \citet{heavens/matarrese/verde:1998} computed the redshift space power spectrum for biased tracers in standard perturbation theory.  This approach is complicated by the need to introduce a smoothing scale by which to define local Eulerian biasing, and \citet{roth/porciani:2011} found that this biasing scheme is not very accurate compared with halos in $N$-body simulations.  Finally, standard perturbation theory cannot be used to predict the behaviour of the correlation function because $P^{1-loop}_{SPT}(k)$ diverges for large $k$, which contribute to $\xi(r)$ even for large $r$ \citep[see discussion in Section IIIB of][]{Mat08a}.

Recently \citet{Mat08a,Mat08b} have addressed both of these shortcomings of SPT by developing a Lagrangian perturbation theory (LPT) description including non-linear local {\em Lagrangian} bias in redshift space which does not require the introduction of a smoothing scale, and is well-behaved at large $k$ for $\xi(r)$ to be computed.  In Fig.~\ref{fig:comparematsu}, we compare the prediction from this theory to our $N$-body simulation results for halo clustering for three different halo masses given in Table \ref{table:halos}.  The bottom right panel shows good agreement for both the monopole and quadrupole on BAO scales \citep[see also][]{PadWhiCoh09,NohWhiPad09}.  The upper left panel shows reasonable agreement for $\xi_0$ between the theory and simulations.  However, the theory predictions for $\xi_2$ and $\xi_4$ do not fit the quasilinear scales of interest in this paper ($\sim 30-80\,h^{-1}$Mpc), and so a more accurate theory must be developed in order to infer the peculiar velocity field amplitude from halo clustering in redshift space.
\begin{table}
\begin{center}
\begin{tabular}{lllll}
sample & $log(M)$ range & $\bar{b}_{lin}$ & $\bar{b}_{LPT}$ & $\bar{n} \; (h^{-1} {\rm Mpc})^{-3}$\\
  \hline
high & $>$13.387 & 2.67 & 2.79 & $7.55 \times 10^{-5}$\\
low & 12.484 - 12.784 & 1.41 & 1.43 & $4.04 \times 10^{-4}$\\
HOD & - & 1.84 & 1.90 & $3.25 \times 10^{-4}$\\
\end{tabular}
\caption{\label{table:halos} Halo samples from $N$-body simulations.  The first two samples are defined by sharp cuts in FoF halo mass (linking length 0.168), while the third sample contains all halos that host at least one galaxy in the mock catalogues of \citet{white/etal:2011}.  This is equivalent to a catalogue of all mock central galaxies and no satellite galaxies.  The bias values $\bar{b}_{lin}$ and $\bar{b}_{LPT}$ we list are derived by fitting the real space halo correlation function to the linear theory and Lagrangian perturbation theory predictions for $\xi(r)$, including scales $r > 30\,h^{-1}$Mpc.  $\bar{b}_{lin}$ assumes a scale-independent linear bias, while $\bar{b}_{LPT}$ is fit including the second order bias implied from the peak background split (see \citet{Mat08b} for details).}
\end{center}
\end{table}
\subsection{A non-perturbative real-to-redshift space mapping: the scale-dependent Gaussian streaming model}
\label{sec:streaming}

\begin{figure}
\begin{center}
\resizebox{\columnwidth}{!}{\includegraphics{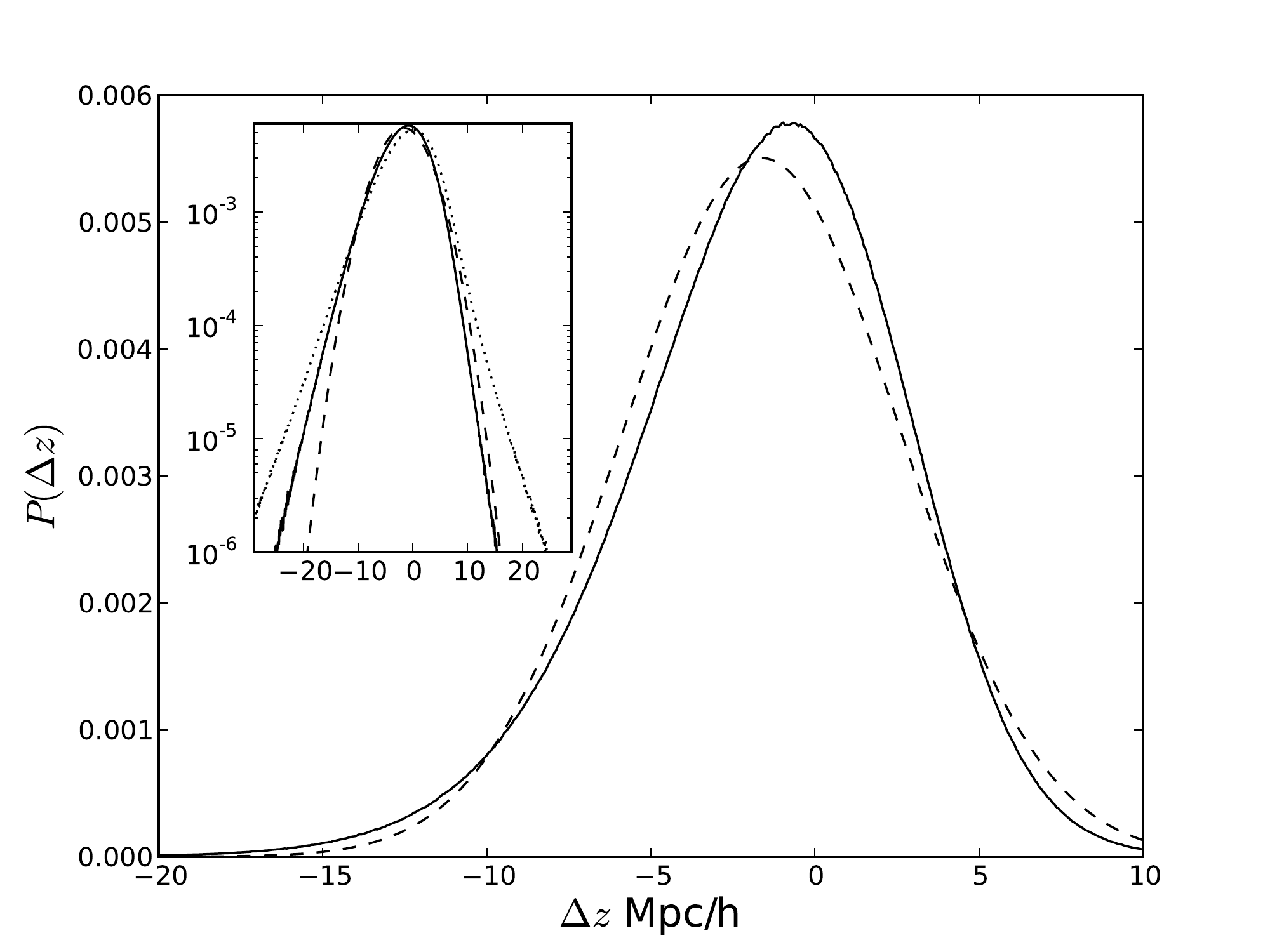}}
\end{center}
\caption{The radial pairwise halo velocity probability distribution function for the HOD halo subsample in Table \ref{table:halos} for pair separations $30\,h^{-1}$Mpc $< r < 31.5\,h^{-1}$Mpc (solid curves) compared with a Gaussian distribution (dashed curves) with the same mean ($-1.6\,h^{-1}$Mpc) and variance $18\,(h^{-1}$Mpc$)^2$.  The normalisation is arbitrary.  While the halo PDF has clear skewness and kurtosis, the PDF for dark matter particles in our simulation has 30 per cent larger variance and exponential tails.  The mean infall between dark matter particles ($-0.9\,h^{-1}$Mpc) is smaller than the more highly biased halo samples.}
\label{fig:vpdf}
\end{figure}
\begin{figure*}
\begin{center}
\resizebox{\textwidth}{!}{\includegraphics{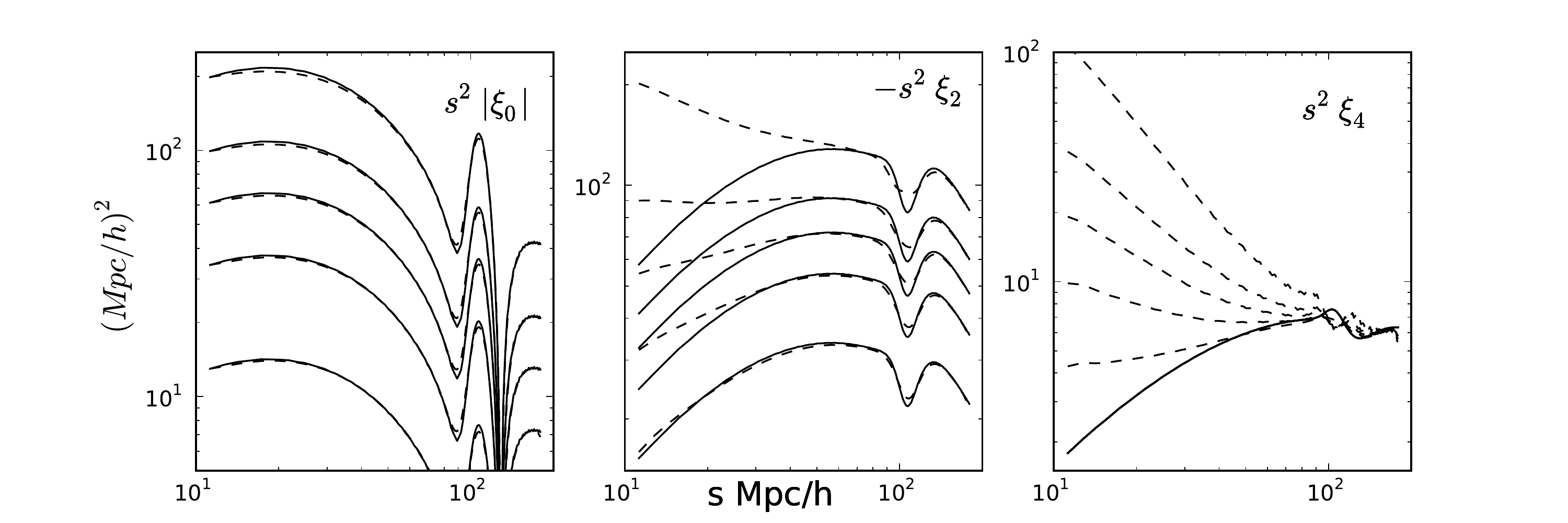}}
\end{center}
\caption{The solid curves show the linear theory Kaiser formula predictions for the Legendre moments of the redshift space correlation function, $\xi_{0,2,4}(s)$ given in Eq.~\ref{eq:ximoments}, where $s$ is the redshift space pair separation for the three different bias values given in Table \ref{table:halos}, as well as $b=1$ and $b=0.5$ for comparison.  The dashed lines show the predictions of the scale-dependent Gaussian streaming model, Eq.~\ref{streamingeqn}, which we evaluated using the linear theory expectations for $\xi^{r}_g(r)$, $v_{12}(r)$, and $\sigma_{12}^2(r, \mu)$.  We see that accounting for the full real-to-redshift space mapping for a Gaussian pairwise velocity probability distribution that agrees with linear theory significantly modifies $\xi_2$ and $\xi_4$ from the Kaiser formula expectation on quasilinear scales, with larger corrections at higher bias.}
\label{fig:streaminglin}
\end{figure*}

As perturbation theory does a good job of describing $P_{\delta \theta}$ and $P_{\theta \theta}$ on the relevant scales \citep[][see also our section \ref{sec:PT}]{carlson/white/padmanabhan:2009} we conjecture that the failure of perturbation theory descriptions of the redshift space power spectra can be traced to the inaccuracy of a perturbative description of the real-to-redshift space mapping \citep{SCF99}.
\cite{Fis95} derived the exact result for the redshift space correlation function in the case where both density and velocity fields are Gaussian and related to one another as in linear theory (our Eq.~\ref{eq:exactgaussian}).  Thereafter, several authors \citep{Bha01,Sco04,ShaLew08} showed that there are significant differences between the exact result and the Kaiser limit, Eq.~\ref{pkkaiser}, which can be traced to additional assumptions inherent in the Kaiser derivation.  However, we found that this exact mapping for Gaussian fields does not improve agreement with $N$-body simulation results for halo clustering in redshift space.

A principle object of interest in the study of redshift space clustering statistics is the pairwise velocity probability distribution function (PDF) $\mathcal{P}(v_z, {\bf r})$, i.e., the probability that a pair with real space separation ${\bf r}$ has relative LOS velocity $v_z$.  Even for the exact result in the Gaussian case (Eq.~\ref{eq:exactgaussian}), the corresponding pairwise velocity PDF is non-Gaussian; however, near its peak it can be approximated by a Gaussian centred on $\mu v_{12}(r)$ \citep{Sco04}.  By resumming the mean infall velocity term ($\propto \mu v_{12}(r)$) from Eq.~\ref{eq:exactgaussian} into the exponential, we recover a streaming model expression that agrees to linear order with the Gaussian case, and for which the corresponding pairwise velocity PDF's mean and dispersion are correct:
\begin{equation}
  1+\xi^{s}_g(r_{\sigma},r_{\pi}) = \int \left[1+\xi^{r}_g(r)\right]
  e^{-[r_\pi - y - \mu v_{12}(r)]^2/2\sigma_{12}^2(r,\mu)} \frac{dy}{\sqrt{2\pi\sigma^2_{12}(r,\mu)}} \label{streamingeqn}.
\end{equation}
Here $r = r_{\sigma}^2 + y^2$ is the pair separation in real space and $\mu = y/r$.  This model sums over all pairs with real space  LOS separation $y$ that are mapped to redshift space LOS separation $r_{\pi}$ with a Gaussian probability distribution whose mean and dispersion depend on both scale $r$ and angle with the LOS, $\mu$.  Note that our approach is not mathematically rigorous, but is rather an attempt to capture the most important non-perturbative effects.  It can be thought of as a generalisation of models such as Eq.~\ref{eq:pkgauss} to include the scale dependence of the exponential terms.

Non-linear evolution causes the dark matter pairwise velocity PDF to be significantly non-Gaussian on all scales; Eq.~\ref{streamingeqn} neglects this.  In Fig.~\ref{fig:vpdf} we compare the halo pairwise velocity PDF (solid) with that of the dark matter (dotted) for pairs separated by $30\,h^{-1}$Mpc $< r < 31.5\,h^{-1}$Mpc.  The dashed lines show a Gaussian PDF with the same mean and variance as the halo PDF measured from the simulations.  The PDF for the dark matter has a lower mean infall (as expected in linear theory).  The dark matter PDF has 30 per cent larger variance than the halos, and clear exponential tails at large velocities, as one would expect given that intrahalo virial velocities contribute only to the dark matter PDF.  In contrast, the variance of the halo PDF is close to the expectation from linear theory.  While skewness and kurtosis are clearly evident in the halo pairwise velocity separations at $r = 30\,h^{-1}$Mpc, if their scale dependence is smooth, $\xi^s$ may be relatively immune to them.  Therefore, we will explore the accuracy of Eq.~\ref{streamingeqn} for the redshift-space clustering of halos in the rest of this paper, and show that for many of the statistics of interest, this approximation is adequate on scales of tens of $h^{-1}$Mpc.  We also note that \citet{tinker:2007} characterises the skewness and kurtosis of of the pairwise velocity PDF, and develops a physical model that describes them based on how a halo's environment depends on its mass.  However, the improved accuracy of his model on smaller scales than are of interest in this paper comes at the cost of much greater model complexity than Eq.~\ref{streamingeqn}.  His fig. 5 demonstrates that the relative deviations from Gaussianity quickly decline as a function of pair separation for $r > 10\,h^{-1}$Mpc, so we expect Eq.~\ref{streamingeqn} to improve in accuracy as we move to larger scales. 

Allowing for the non-perturbative real-to-redshift space mapping given in Eq.~\ref{streamingeqn} alters of the redshift space correlation function substantially in the quasilinear scales of interest in this paper
($\sim 30-80\,h^{-1}$Mpc), even when we insert the linear theory predictions for $\xi^r_g$, $v_{12}(r)$, and $\sigma^2_{\parallel/\perp}(r)$ (Equations \ref{linearinfall} to \ref{vdisptot}); we show this in Fig.~\ref{fig:streaminglin}.  The mapping both smooths the BAO feature \citep[but see also][]{tian/etal:2011} and enhances $\xi_{2,4}$ on quasilinear scales, while only slightly suppressing $\xi_0$.  Clearly, the amplitude of the correction terms to the Kaiser prediction depend strongly on bias; we will study the bias dependence in more detail in Section \ref{biasdep}.

\section{Setting the bar: target model precision}
\label{fisher}
\begin{table}
\begin{center}
\begin{tabular}{llll}
z & $f\sigma_8$ & survey \\
  \hline
0.17 & $0.51 \pm 0.06$ & 2dFGRS \\
0.77 & $0.49 \pm 0.18$ & VVDS \\
1.5 & $0.72 \pm 0.15$ & 2SLAQ \\
0.25 & $0.39 \pm 0.05$ & SDSS \\
0.37 & $0.43 \pm 0.04$ & SDSS \\
0.22 & $ 0.42 \pm 0.07 $ & WiggleZ \\
0.41 & $ 0.45 \pm 0.04 $ & WiggleZ \\
0.6 & $ 0.43 \pm 0.04 $ & WiggleZ \\
0.78 & $ 0.38 \pm 0.04 $ & WiggleZ \\
\end{tabular}
\caption{\label{table:fs8constraints} Constraints on $f\sigma_8$  from \citet{percival04}, \citet{guzzo08}, \citet{Ang08}, \citet{samushia/percival/raccanelli:2011}, and \citet{blake/etal:2011}, respectively.}
\end{center}
\end{table}

Before entering a detailed study of the accuracy with which various models reproduce the clustering of halos in $N$-body simulations, we consider the level of model accuracy desired in the near-term future.  One aspect of this problem is the raw statistical power of current and upcoming surveys; for concreteness we focus on the ongoing Baryon Oscillation Spectroscopic Survey \citep[BOSS;][]{schlegel/white/eisenstein:2009,eisenstein/etal:2011}.  The other aspect to consider is the level at which competing theories differ in redshift-space distortion observables, and on what scales the differences are largest.

For any expansion history $H(z)$ and negligible dark energy clustering, general relativity makes a unique prediction for the growth of structure as probed by redshift space distortions; alternative theories of gravity and/or dark energy could potentially alter the predicted velocity field amplitude by $\sim 10$ per cent level
\citep[e.g.][but unfortunately there is no generic prediction for the level of deviation from $\Lambda$CDM+GR for alternative models]{Song08b}.  Furthermore, combining precise measurements of peculiar velocities using redshift space distortions with lensing measurements provides an additional, powerful test of general relativity \citep{Zhang07,Song08a, reyes/etal:2010}.  Therefore, extracting constraints from quasilinear scales where lensing can be measured with high signal-to-noise is desirable.

In Table \ref{table:fs8constraints} we compile measurements of $f\sigma_8$ from galaxy redshift surveys; see \citet{Song08b} for the details on how these numbers were inferred from the reported constraints.  In particular, recent analyses of the SDSS LRG sample \citep{samushia/percival/raccanelli:2011} and from the WiggleZ Dark Energy Survey \citep{blake/etal:2011} provide constraints on the growth of structure at the $\sim 7$ per cent and $\sim 5$ per cent level, in the limit that each redshift bin can be treated independently.
As we show in the next section, the Baryon Oscillation Spectroscopic Survey \citep[BOSS;][]{schlegel/white/eisenstein:2009,eisenstein/etal:2011} will measure the peculiar velocity field at the few percent level.  Thus, current and near-term observational constraints demand better precision than is currently available from analytic models.

\subsection{Fisher Matrix Analysis in Configuration Space}

\begin{figure}
  \centering
  \resizebox{\columnwidth}{!}
{\includegraphics{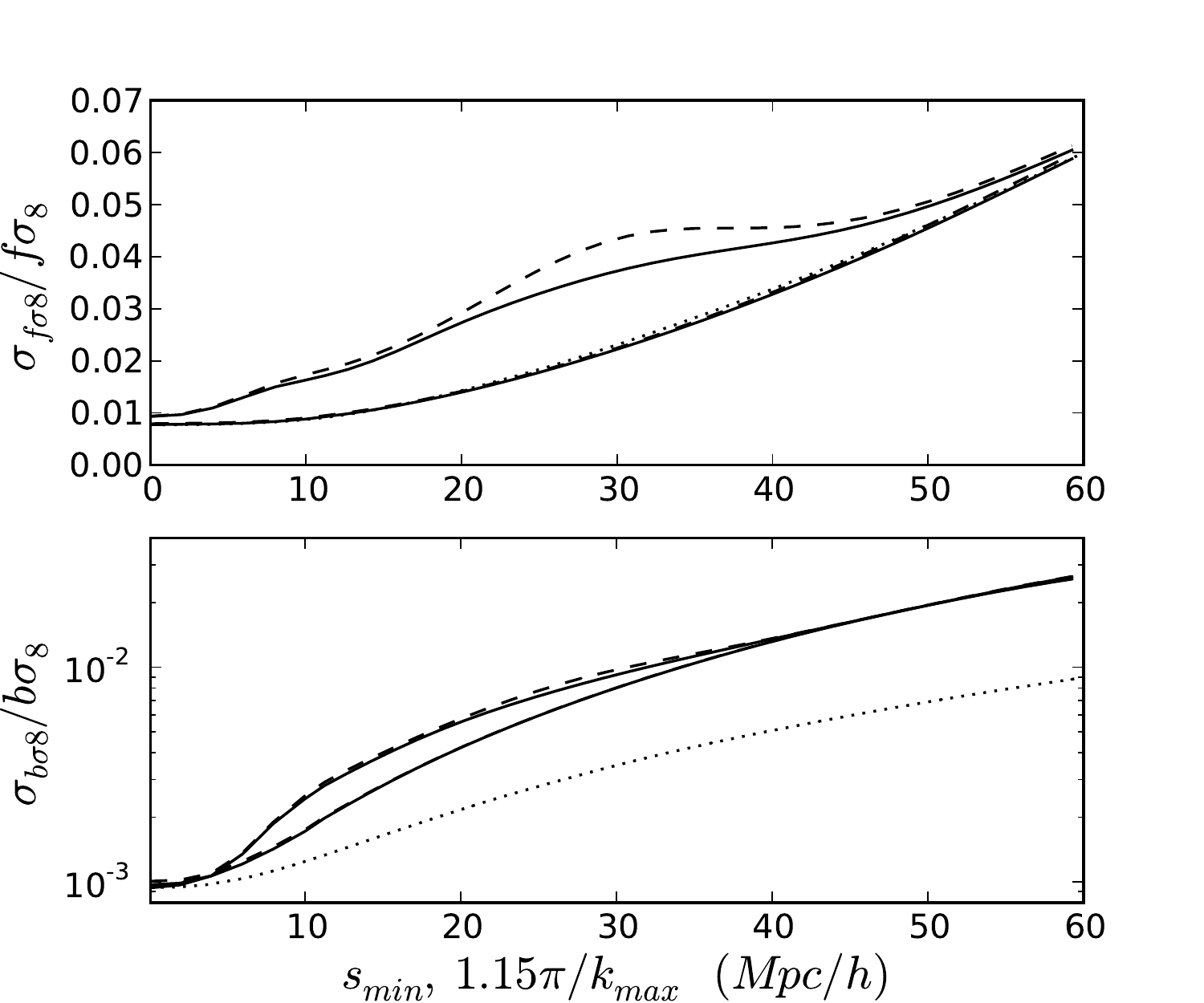}}
  \caption{\label{fig:fisher1}  Fractional error on $f\sigma_8$ and $b\sigma_8$ as a function of the minimum configuration space separation $s_{min}$ used in the analysis, assuming a covariance matrix given by the usual Gaussian cosmic variance term and Poisson sampling variance term, with $b=2$, $V_{survey} = 5 \,h^{-3}$Gpc$^{3}$, and $\bar{n} = 3.0 \times 10^{-4} \,h^{3}$Mpc$^{-3}$.  We use the model in Eq.~\ref{eq:pkgauss}, and consider two cases.  In the first, we marginalise over the value of $\sigma$, which controls the small scale damping.  In the second case, we assume $\sigma$ is perfectly known; in both the fiducial value of $\sigma$ is $3.5\,h^{-1}$Mpc.  Marginalisation over $\sigma$ increases the error on $b\sigma_8$ and $f\sigma_8$ (upper two curves in both panels).  The dashed curves show constraints when only $\xi_{0,2}$ are used; the solid curves include $\xi_4$ as well.  In the case where $\sigma$ is known (lower curves), $\ell_{max} = 2$ and 4 are indistinguishable.  Finally, 
we compare the configuration space results to an analysis in Fourier space \citep{WhiSonPer09}, where a sharp cut-off in wavenumber is imposed, shown as dotted curves as a function of $1.15\pi/k_{max}$ and also assuming $\sigma$ is known.}
\end{figure}

In this section we present a calculation of the linear theory constraints on $b\sigma_8$, $f\sigma_8$, and $\sigma$, using Eq.~\ref{eq:pkgauss} as our underlying model, and assuming perfect knowledge of $P^r_m(k)$; the same calculation was done in \citet{WhiSonPer09} in Fourier space.

Throughout this section, we work at $z=0.55$ where $f=0.74$ in our fiducial cosmology.  Our fiducial value for $\sigma$ is $300\,{\rm km}\,{\rm s}^{-1}=3.5\,h^{-1}$Mpc.  Fig.~\ref{fig:fisher1} compares the relative error on $f\sigma_8$ and $b\sigma_8$ as a function of the minimum redshift space pair separation $s_{\rm min}$ used in the analysis; we assume $s_{\rm max} = 180\,h^{-1}$Mpc.  In the figure, solid lines show constraints including $\xi_{0,2,4}$, while dashed lines include only $\xi_0$ and $\xi_2$.  We consider two cases: one in which $\sigma$ is marginalised over (upper two curves in both panels), and the other where $\sigma$ is assumed known (lower three curves in both panels).
The parameter $\sigma$ in Eq.~\ref{eq:pkgauss} traditionally represents the effect of redshift errors and small scale velocity dispersions (fingers-of-god), but has also been used to absorb some of the theoretical uncertainty in the redshift space power spectrum \citep[e.g.,][]{PerWhi08,TarNisSai10}.  If $\sigma$ can be constrained by better modelling of redshift space halo clustering in the quasi-linear regime and/or complementary observables sensitive the small-scale galaxy velocity distribution, Fig.~\ref{fig:fisher1} indicates a large sensitivity gain to the large scale peculiar velocity field amplitude for $s_{\rm min} \sim 30\,h^{-1}$Mpc; in this paper we will focus only on the former approach.

Except in the case of unknown $\sigma$, in the range $20\,h^{-1}$Mpc$ < s_{\rm min} < 40\,h^{-1}$Mpc very nearly all of the information on $f\sigma_8$ is contained in $\xi_{0}$ and $\xi_{2}$.  We therefore focus our modelling efforts on those two moments, though accurate modelling of $\xi_4$ will be important if the geometric factors $D_A$ and $H$ are also being fitted \citep{TarSaiNis11,kazin/sanchez/blanton:2011}.

Finally, in the case of fixed, perfectly known $\sigma$, we can compare the constraints in configuration space to those in Fourier space using the public code released with \citet{WhiSonPer09}, shown as dotted curves as a function of $1.15\,\pi/k_{\rm max}$, which provides an empirical mapping between $s_{\rm min}$ and $k_{\rm max}$ that gives nearly identical errors on $f\sigma_8$ over the range of scales we consider.  Reassuringly, the same fractional errors are recovered in both configuration and Fourier space when all scales are included.  While the relatively smaller errors on $b\sigma_8$ in Fourier space indicate that there is more information on the shape of the power spectrum at fixed non-linear scale $s_{\rm min} \sim \pi/k_{\rm max}$, the present analysis ignores the fact that shot noise will contribute at all $k$ to the power spectrum, and cannot be assumed Poissonian at this level of accuracy for halos or galaxies \citep[e.g.][]{cooray/sheth:2002,smith07,hamaus/etal:2010}.

\section{Comparing the Scale-dependent Gaussian streaming model for halos with $N$-body simulations}
\label{simxis}

\subsection{Simulations}

To assess the accuracy of our analytic model we compare it to halo correlation functions derived from the 20 $N$-body simulations of \citet{white/etal:2011} which have $L_{\rm box}=1.5\,h^{-1}$Gpc and particle masses $m_p = 7.6 \times 10^{10} h^{-1} M_{\odot}$ \citep[see][for more details]{white/etal:2011}.  The simulations are of a $\Lambda$CDM cosmology with $\Omega_m = 0.274$, $\Omega_{\Lambda} = 0.726$, $h=0.7$, $n=0.95$, and $\sigma_8 = 0.8$.  In this section we consider only a single snapshot at $z=0.55$.  The halo catalogues are generated by the Friends of Friends (FoF) algorithm \citep{davis/etal:1985} with linking length 0.168 times the mean inter-particle spacing.  The principal purpose of this investigation is to generate an accurate model of redshift space distortions as a function of halo bias.  Throughout this section we will compare models with three distinct halo samples; the same level of agreement was found for all mass bins we investigated.  The detailed properties of the samples are listed in Table \ref{table:halos} -- the `high' and `low' samples impose sharp halo mass cuts, while the `HOD' sample includes all halos hosting a mock galaxy in the catalogues used in \citet{white/etal:2011}.  This catalogue is different from the mock galaxy catalogue only in that the ``satellite'' galaxies are excluded from the sample.  The purpose of this sample is to demonstrate that a sample including a broad range of halo masses can still be described by our model, bringing us one step closer to modelling what we observe: the {\em galaxy} density field.

\subsection{Results}

\begin{figure}
  \centering
  \resizebox{\columnwidth}{!}
{\includegraphics{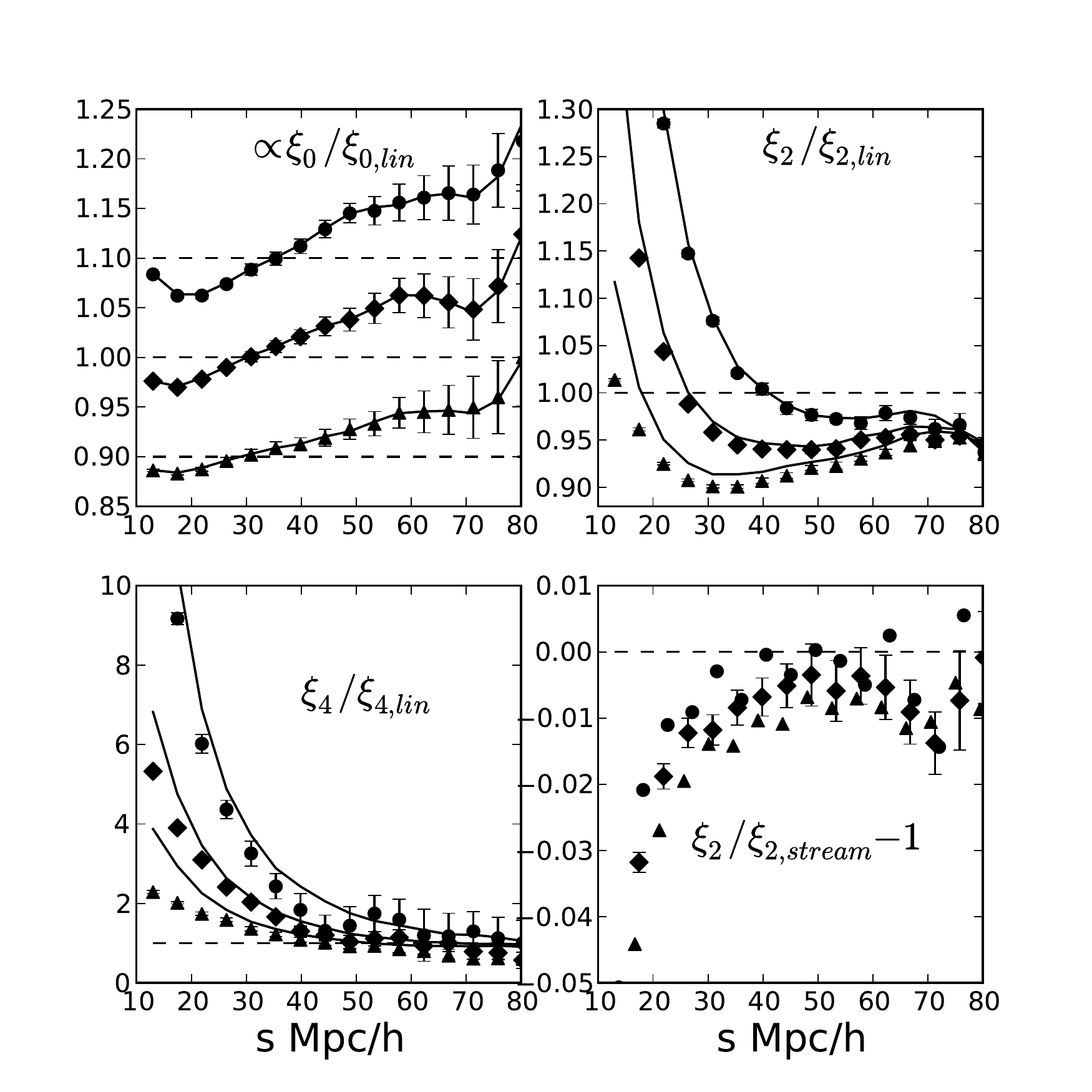}}
  \caption{\label{fig:streaming1}  Redshift space Legendre moments $\xi_{0,2,4}$ as a function of redshift space separation $s$ divided by the linear theory expectations.  Points with errors show measurements from $N$-body simulations for the high (circles), low (triangles), and HOD (diamonds) samples described in Table \ref{table:halos}.  In the upper left panel we offset the ``high'' and ``low'' halo sample by $\pm 10$ per cent, respectively, for clarity.  Black solid curves show the result of the scale-dependent streaming model, Eq.~\ref{streamingeqn}, in the case where $\xi^{r}_h(r)$, $v_{12}(r)$, and $\sigma_{12}^2(r, \mu)$ have been measured directly from the $N$-body simulation {\em real} space clustering and velocity statistics for the same halos.  This comparison tests the approximations in the form of Eq.~\ref{streamingeqn}, supposing we had a perfect theory for the input real space statistics.
The bottom right panel shows the fractional error of the streaming model on $\xi_2$ as a function of scale.  For all halo mass bins we studied, the streaming model is accurate at the 2 per cent level (or better) at $s=25 \,h^{-1}$Mpc.}
\end{figure}

In this section we test the approximations entering the scale-dependent Gaussian streaming model, Eq.~\ref{streamingeqn}, for halos in $N$-body simulations.  We do so by assuming we have a perfect model for all the real-space quantities entering Eq.~\ref{streamingeqn}: the halo correlation function $\xi^{r}_{h}(r)$, the mean pairwise infall velocity as a function of separation, $v_{12}(r)$, and the pairwise velocity dispersions parallel and perpendicular to the pair separation vector, $\sigma^2_{\parallel,\perp}(r)$.  We simply measure these quantities from our $N$-body simulations and input them into Eq.~\ref{streamingeqn} to generate model curves for Fig.~\ref{fig:streaming1}, shown as black solid curves (we will consider analytic models for these functions in Section \ref{sec:PT}).

Fig.~\ref{fig:streaming1} shows $\xi_{0,2,4}$ for the high (circles), low (triangles), and HOD (diamonds) halo subsamples in Table \ref{table:halos}; these are the same as in Fig.~\ref{fig:comparematsu}.  We again offset the `high' and `low' samples by $\pm 10$ per cent for clarity in the $\xi_0$ panel.  In the lower right we plot the fractional difference between the simulation results and our model for the quadrupole $\xi_2$, highlighting the percent level agreement between the streaming model and the $N$-body results down to $s \sim 25\,h^{-1}$Mpc.  

For clarity we omit the results on BAO scales; the agreement is as good as the \citet{Mat08b} model shown in the bottom right panel of Fig.~\ref{fig:comparematsu}.  On smaller, ``quasi-linear'' scales, we find a drastic improvement over the model of \citet{Mat08b}, which is unsurprising since our model includes real space halo clustering and velocity statistics derived from $N$-body simulations, and treats the non-linear mapping between real and redshift space non-perturbatively.  The streaming model provides a sufficiently accurate description of the monopole $\xi_0$ down to $10\,h^{-1}$Mpc, with no detectable deviation even for this large volume of simulations.  The streaming model is an accurate model for $\xi_2$ at the 1-2 per cent level for redshift space separations greater than $\sim 25\,h^{-1}$Mpc.  However, both the LPT and streaming model predictions are poorer descriptions to the $N$-body results for higher multipole moments.  On scales $s < 80\,h^{-1}$Mpc, $\xi_4$ deviates from the linear theory prediction by O(1).  However, the streaming model has the right dependence on bias and may be sufficiently accurate on some range of scales for upcoming surveys, given the large measurement errors on $\xi_4$.  We are also optimistic that treatment of non-Gaussian small-scale velocities that are uncorrelated with the linear theory pairwise infall velocities we wish to measure may absorb some of the remaining discrepancy.  We reserve this line of study for future work.

As we emphasised earlier, both $P_{\delta \theta}$ and $P_{\theta \theta}$ are suppressed compared with linear theory on relatively large scales ($k \gtrsim 0.03\,h\,{\rm Mpc}^{-1}$), and we expect that suppression to affect redshift space clustering on similar scales.  Even at $s = 50\,h^{-1}$Mpc, $\xi_2$ is suppressed by 2, 6, and 8 per cent compared to the linear theory expectation, in line with the findings of \citet{okumura/jing:2011}.  Comparison with Fig.~\ref{fig:streaminglin} shows that on the scales of interest ($\sim 30-80\,h^{-1}$Mpc), both non-linear corrections to the velocity field and the non-linear mapping from real to redshift space are relevant to predicting $\xi_{2}$, at a level comparable to the latest $\sim 7$ per cent peculiar velocity field constraints \citep{samushia/percival/raccanelli:2011,blake/etal:2011}, and certainly at the level of constraints projected for BOSS in Fig.~\ref{fig:fisher1}.

\section{Components of the Scale-dependent Gaussian streaming model: Perturbation Theory}
\label{ptsec}

In this section, we consider perturbation theory models for the components of the scale-dependent Gaussian streaming model in pursuit of an analytic model for the redshift space correlation function of biased tracers.  While our model for the real space halo correlation function $\xi^{r}_h(r)$ includes non-linear Lagrangian bias, our perturbation theory calculations for halo infall velocities and dispersions were carried out in standard perturbation theory, where considering non-linear biasing requires the definition of a smoothing scale \citep{heavens/matarrese/verde:1998}.  \citet{roth/porciani:2011} also found that the Eulerian local biasing scheme is not very accurate, and it is not equivalent to a Lagrangian biasing scheme \citep{Mat11}.  For these reasons, we perform our velocity calculations for the simplest model of {\em linearly} biased tracers with bias $b$, which we identify with the large scale bias we fit to the real space correlation function using LPT.  

\subsection{Halo real space correlation function, $\xi^{r}_h(r)$}
\label{sec:PT}

\begin{figure}
  \centering
  \resizebox{\columnwidth}{!}
{\includegraphics{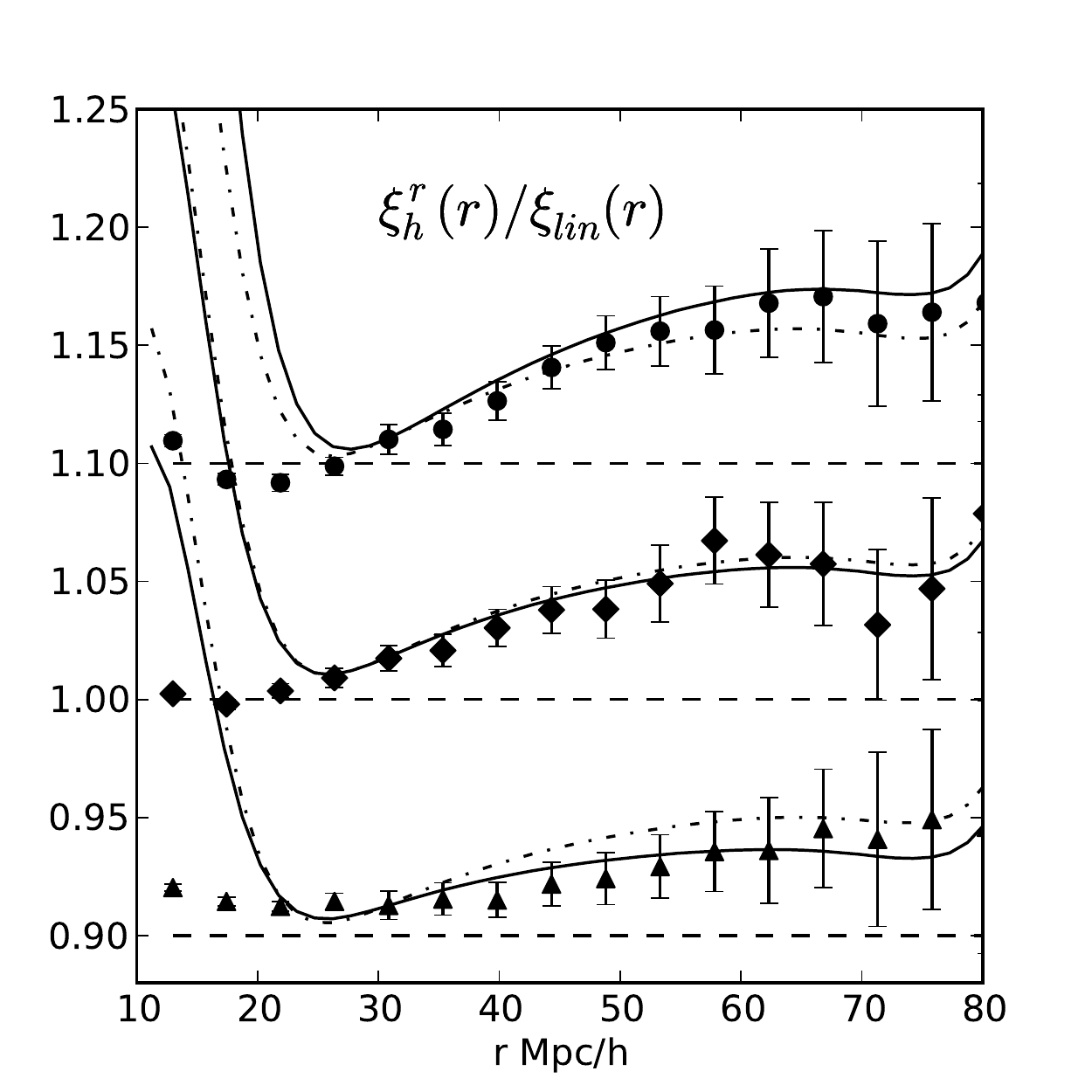}}
  \caption{\label{fig:xirealPTfit}  Real space correlation functions for the halo samples in Table \ref{table:halos}.  The high (low) bias bins are offset by $\pm 10$ per cent for clarity.  The solid curve shows the prediction of \citet{Mat08b} when we fit for the large scale bias as a free parameter.  For the dashed-dot curves, we show the \citet{Mat08b} prediction when we artificially set $b_2=0$, just to show the order of magnitude of the contribution from non-linear halo bias.  Using the linear theory covariance matrix, we find a good fit for scales $r > 25\,h^{-1}$Mpc to our simulation results, which total $67.5\,h^{-3}$Gpc$^{3}$.  BAO scales are not shown here, but the fit is good on those scales as well.}
\end{figure}

The Lagrangian perturbation theory prescription of \citet{Mat08b} for describing local {\em Lagrangian} biased objects includes both first and second order bias terms, which are related by the peak-background split to the halo mass function.  This theory provides a good description (accurate at the 1 per cent level) on scales $r > 25\,h^{-1}$Mpc.  We show this explicitly for scales $r < 80\,h^{-1}$Mpc in Fig.~\ref{fig:xirealPTfit}, and note that the fit is good on the full range of scales we study ($r < 180\,h^{-1}$Mpc).  If we fit the LPT prediction to $\xi^{r}_h(r)$ for separations
$30 \,h^{-1}$Mpc $<r < 180 \,h^{-1}$Mpc, we find $\chi^2 = 93, 104, 119$ for 99 degrees of freedom and one free parameter (the large-scale halo bias) for the high, low, and HOD halo subsamples, respectively.  Fitting instead to linear theory gives $\chi^2 = 212, 274, 403$.  In both cases we use the standard linear theory with the Poisson sampling assumption to derive the covariance matrix $\left\langle\Delta \xi(r_i) \Delta \xi(r_j)\right\rangle$.  To illustrate the amplitude of the second order bias corrections for the halos of interest, we also plot the LPT prediction when $b_2$ is artificially set to zero (dashed-dot curve) compared with the LPT prediction including nonzero $b_2$ (solid curve); the second order bias contribution to $\xi^{r}_h$ is quite small for the linear halo bias values we consider.

\subsection{Mean halo infall velocities, $v_{12}(r)$}

\begin{figure}
  \centering
  \resizebox{\columnwidth}{!}
{\includegraphics{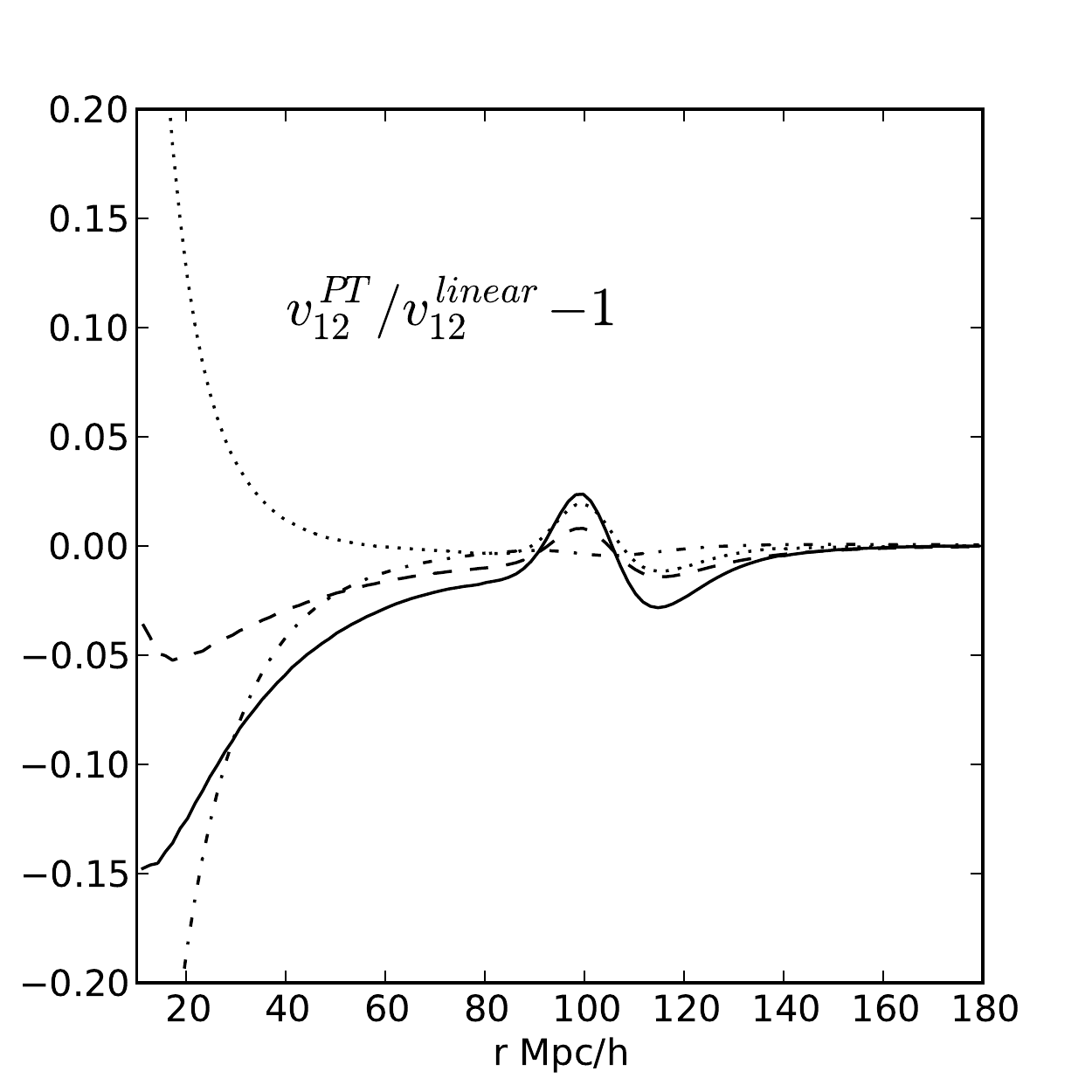}}
\caption{\label{fig:PTtot}  The three components of the next-to-leading order correction to the linear pairwise infall, $v^{\rm lin}_{12}(r)$; each term scales differently with bias.  The solid curve shows the total prediction for $b=2$ halos at $z=0.55$.  The dashed curve shows the contribution from the $\left\langle\delta_i v_j\right\rangle$ terms in Eq.~\ref{vinfallPT} (i.e., those originating from corrections to $P_{\delta \theta}$), the dotted curve shows the contribution from $b \left\langle\delta_i \delta_j v_k\right\rangle$ terms (i.e., those originating from $B_{\delta \delta \theta}$), and the dash-dot curve shows the pair-weighting correction contribution, $1/[1+ b^2\xi_m^r(r)]-1$.  Note that the $P_{\delta \theta}$ correction terms never dominate the total correction, so we should generically expect that models neglecting the three point and pair-weighting corrections to be poor models of the redshift space power spectrum for halos (e.g., Eq.~\ref{scoccPT}).}
\end{figure}

A mean (pairwise halo) velocity along the pair separation vector arises from the correlation of the density field with the velocity field:
\begin{equation}
  v_{12}(r) \hat{r} = \frac{\left\langle [1+b \delta({\bf x})] [1+ b \delta({\bf x}+{\bf r})]
  [{\bf v}({\bf x}+{\bf r}) - {\bf v}({\bf x})] \right\rangle}
  {\left\langle [1+b \delta({\bf x})] [1+ b \delta({\bf x}+{\bf r})] \right\rangle}
\label{eq:vinfall}
\end{equation}
where $b$ is the linear halo bias.  In perturbation theory, the density and velocity fields are written as a sum of terms ($\delta = \delta_1 + \delta_2 + \delta_3 + ...$), with the subscript denoting the order of their dependence on the linear density field, $\delta_1({\bf k})$.  Up to fourth order in $\delta_1({\bf k})$, there are three distinct corrections to the linear theory expectation $v_{12}(r)$ given in Eq.~\ref{linearinfall}, each with a different dependence on bias:
\begin{eqnarray}
\left[1+b^2\xi_m^r(r)\right] v^{PT}_{12}(r) \hat{r} = 2b\left\langle\delta_1({\bf x}) {\bf v}_1({\bf x}+{\bf r})\right\rangle + \nonumber \\
2b\sum_{i>0} \left\langle\delta_i({\bf x}) {\bf v}_{4-i}({\bf x}+{\bf r})\right\rangle + 2b^2 \sum_{i,j>0} \left\langle\delta_i({\bf x}) \delta_j({\bf x}+{\bf r}) {\bf v}_{4-i-j}({\bf x}+{\bf r})\right\rangle \label{ptvinfall} \label{vinfallPT}.
\end{eqnarray}
The three $\left\langle\delta_i v_{4-i}\right\rangle$ terms arise from the perturbation theory corrections to $P_{\delta \theta}$, and the bias dependence is the same as the linear theory term, $\left\langle\delta_1 v_1\right\rangle$.  The terms from three-point correlations $\left\langle\delta_i \delta_j v_{4-i-j}\right\rangle$ scale with $b^2$, so their contribution will be larger for more highly biased tracers.  Note that these terms are exactly the ones evaluated in Appendix B of \cite{tang/kayo/takada:2011}.  We provide explicit expressions for all of these terms in Appendix \ref{ptcalcs}.  Finally, the pair-weighting correction, $1/[1+b^2\xi^r_m(r)]$, will be larger at a given scale for more biased objects.  The relative contribution for these three corrections is shown in Fig.~\ref{fig:PTtot} for $b=2$.  At least for $b=2$ of interest to BOSS, the two-point corrections from $P_{\delta\theta}$ never dominates, so only including the two-point corrections (as in Eq.~\ref{scoccPT}) will be a poor model for the redshift space power spectrum; we should expect important contributions from the bispectrum as well \citep{Sco04,TarNisSai10,tang/kayo/takada:2011}.

\begin{figure*}
  \centering
  \resizebox{7.5in}{!}
{\includegraphics{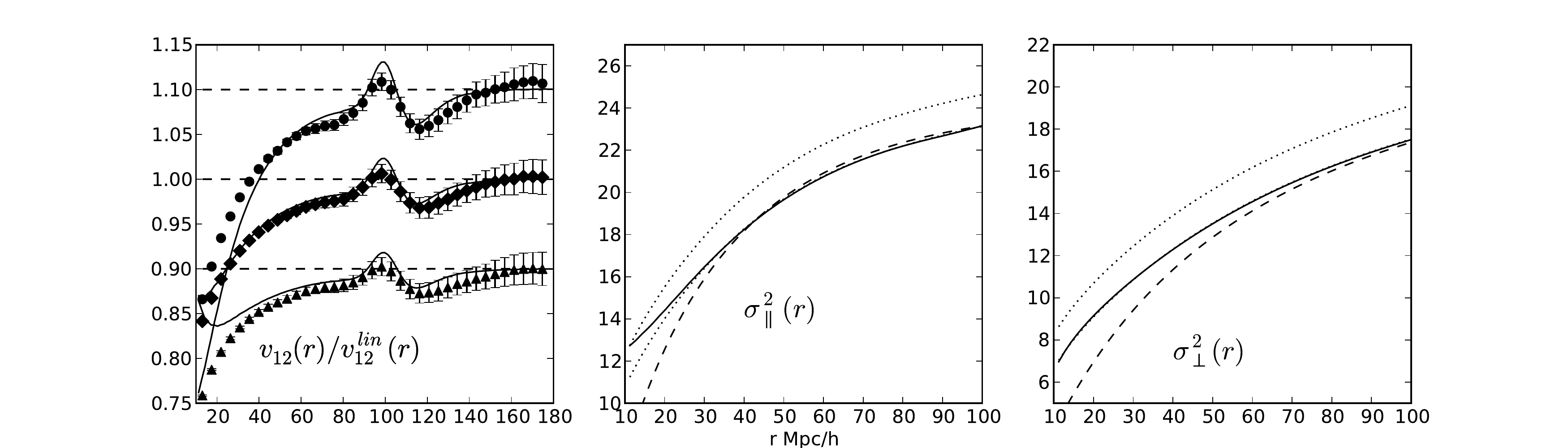}}
\caption{\label{fig:vstats}  {\em Left panel:} Pairwise halo infall velocities for the halo subsamples listed in Table \protect\ref{table:halos} divided by the linear theory expectation (Eq.~\ref{linearinfall}); the high (circles) and low (triangles) bias bins are offset by $\pm 10$ per cent for clarity relative to the HOD sample (diamonds).  Note that the expected $v_{12}^{\rm lin}(r)$ depends on halo bias; we use the best fit LPT bias given in Table \ref{table:halos}.  The solid curves show the prediction for linearly-biased tracers from our perturbation theory calculation, Eq.~\ref{ptvinfall}.  {\em Middle and right panels}: Pairwise halo velocity dispersions measured from the simulations are shown as dotted curves for the directions along and perpendicular to the pair separation vector.  For $\sigma_{\parallel}^2(r)$ we subtract the square of the mean infall $v^2_{12}(r)$ to get the dispersion about the mean, and account for this accordingly in the perturbation theory prediction.  On large scales there is a small offset from the linear theory predictions (dashed curves).  We shift the measured dispersions by a constant value (lower dotted curves) to compare the scale-dependence with both linear theory (dashed) and perturbation theory (solid; also shifted to agree with linear theory at $r=180\,h^{-1}$Mpc).  For $\sigma_{\perp}^2(r)$, the agreement between the perturbation theory calculation and the halo dispersions in simulations is so good that the curves are difficult to distinguish in the figure.}
\end{figure*}

In the left panel of Fig.~\ref{fig:vstats} we compare the deviations from linear theory infall velocity predictions measured from our simulations to our perturbation theory calculation.  The expected $v_{12}(r)$ depends on the halo bias, and for this we use the first order bias deduced from fitting the real-space halo clustering to the LPT model of \citet{Mat08b}; Table \ref{table:halos} indicates that the best fit LPT bias can differ at the few percent level from the best fit linear bias.  At the percent level, Fig.~\ref{fig:vstats} shows that the LPT bias predicts the correct infall velocity amplitude on the largest scales.  This confirms the common assumption in the literature that ``velocity bias'' is small, at least for halos in the bias range we have studied.  Perturbation theory provides a relatively good description of the departure from linear theory.  The difference between the simulations and perturbation theory depend on halo bias and agree best for the HOD halo subsample, which Fig.~\ref{fig:xirealPTfit} indicates is the sample with the smallest second-order bias.  We note that there is good theoretical motivation to expect the bias relevant to the matter-velocity cross-correlation to differ from the one inferred from clustering, and have scale-dependence \citep{desjacques/sheth:2010}.  While including a more complicated biasing model may improve agreement with the pairwise velocity statistics of halos, our results suggest that these corrections are small on the scales of interest here.

\subsection{Halo velocity dispersions, $\sigma^2_{12}(r, \mu^2)$}

Analogous to Eq.~\ref{vinfallPT}, we compute the pair-weighted velocity dispersion.  The result depends only on $\mu^2 = \cos^2(\phi_{\ell r})$, where $\phi_{\ell r}$ is the angle between the LOS and the pair separation vector.  As in Eq.~\ref{vdisptot}, we present results for the velocity dispersion perpendicular and parallel to the LOS, which can be combined to give the dispersion as a function of $(r, \mu^2)$.
\begin{equation}
\sigma^2_{12}(r, \mu^2) = \frac{\left\langle (1+b \delta({\bf x})) (1+ b \delta({\bf x}+{\bf r})) (v^{\ell}({\bf x}+{\bf r}) - v^{\ell}({\bf x}))^2 \right\rangle}{\left\langle (1+b \delta({\bf x})) (1+ b \delta({\bf x}+{\bf r})) \right\rangle} \label{eq:vdisp}
\end{equation}
We again separate the terms by their bias and scale dependence, and provide explicit expressions in Appendix \ref{ptcalcs}:
\begin{eqnarray}
\left[1+b^2 \xi(r)\right] \sigma^2_{12}(r, \mu^2) = \nonumber \\
2\left(\left\langle(v^{\ell}({\bf x}))^2- v^{\ell}(\bf{x}) v^{\ell}({\bf x}+{\bf r})  \right\rangle\right) + \label{v2r1} \\
2b\left\langle\delta({\bf x}) (v^{\ell}({\bf x}))^2\right\rangle + \label{v2r2}\\
2b\left(\left\langle\delta({\bf x}) (v^{\ell}({\bf x}+{\bf r}))^2 \right\rangle - 2 \left\langle\delta({\bf x}) v^{\ell}({\bf x}) v^{\ell}({\bf x}+{\bf r}) \right\rangle\right) + \label{v2r4} \\
+2b^2 \left[\left\langle\delta({\bf x})\delta({\bf x}+{\bf r}) (v^{\ell}({\bf x}))^2 \right\rangle -\left\langle\delta({\bf x})\delta({\bf x}+{\bf r}) v^{\ell}({\bf x}) v^{\ell}({\bf x}+{\bf r})\right\rangle\right]. \label{v2r6} 
\end{eqnarray}
The higher order terms in line \ref{v2r1} can be accounted for in the form of Equations \ref{psiperp} and \ref{psiparallel}, but replacing $P_m^r(k)$ with the perturbation theory result for $P_{\theta \theta}(k)$.  The term in line \ref{v2r2} evaluates to a constant, which for our fiducial cosmological parameters is $13.24b f^2$ ($h^{-1}$Mpc)${}^2$.  While the perturbation theory calculation overestimates the amplitude of the effect, Fig.~\ref{fig:vstats} does show a large scale offset between the linear theory velocity dispersions and those measured for halos in our simulations; the offset for our HOD sample is $\sim 1.5\,(h^{-1}{\rm Mpc})^2$.  There is a slight dependence on halo mass, with the high mass sample offset $\sim 1.0\,(h^{-1}{\rm Mpc})^2$ and the low mass sample offset $\sim 2.2\,(h^{-1}{\rm Mpc})^2$.  The offsets for the parallel and perpendicular components are in agreement with each other at the level of $0.1\,(h^{-1}{\rm Mpc})^2$.  In a future paper we expect to accommodate this offset into our theory along with other small-scale isotropic dispersions due to redshift errors and fingers-of-god from satellite galaxies.  However, we point out that for halos, the small-scale, incoherent velocity dispersion is much smaller than the linear theory dispersion $\sigma_v^2 = 21\,(h^{-1}{\rm Mpc})^2$ that appears in Eq.~\ref{scoccPT}.

Line \ref{v2r4} has many terms that depend on $B_{\delta \theta \theta}$; we evaluate them explicitly in the Appendix \ref{ptcalcs}.  Finally, line \ref{v2r6} reduces to 
\begin{equation}
b^2 \xi^r_m(r) \sigma_{12,{\rm lin}}^2(r,\mu^2) + \frac{1}{2} v_{12,{\rm lin}}^2(r) \mu^2 \label{argh}
\end{equation}
This shows that the pair-weighting factor cancels out to leading order, and the dispersion should be increased along the separation vector due to the linear infall.  The middle and right panels of Fig.~\ref{fig:vstats} compare the scale dependence of the HOD halo subsample velocity dispersions parallel and perpendicular to the LOS, respectively.  The upper dotted curves show the dispersions about the mean infall measured from the halos in the simulations.  To compare the scale dependence, we subtract a constant from the measured velocity dispersions (lower dotted curves), forcing agreement with the linear theory expectation (dashed curves) on the largest scales.  The solid curves show the perturbation theory expectation after subtracting the expected mean infall contribution, $\left<v_{12}(r)\right>^2 \mu^2$, and a constant to force agreement on the largest scales.  As the figure shows, the agreement between the scale-dependence of the dispersions predicted from perturbation theory and the simulations is excellent for the HOD subsample.  We find that it is only slightly worse in the radial direction for the other mass bins.

\section{Accuracy of the Perturbation Theory Scale-Dependent Gaussian Streaming Model}
\label{combinesec}

\begin{figure}
  \centering
  \resizebox{\columnwidth}{!}
{\includegraphics{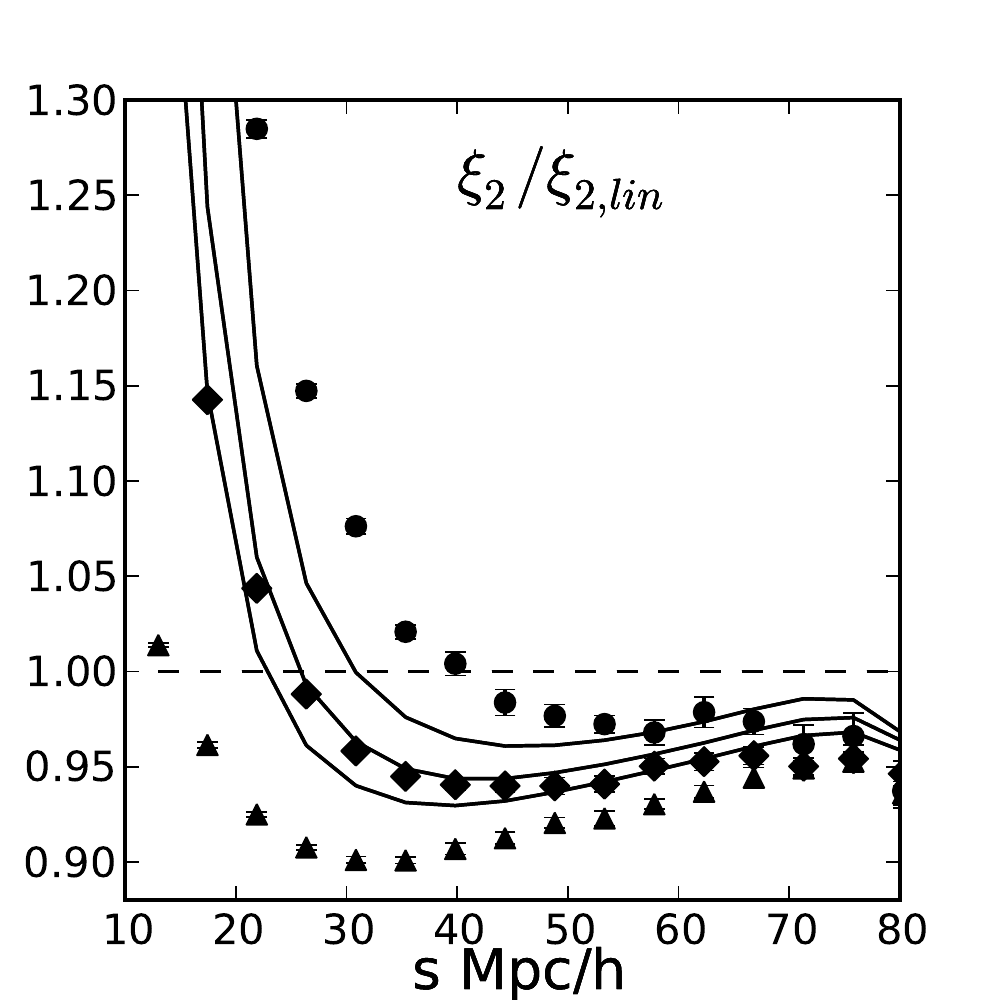}}
\caption{\label{fig:PTstreamxi2}  Inserting the results of Section \ref{sec:PT} into Eq.~\ref{streamingeqn} gives an analytic model for $\xi_{0,2,4}$.  We compare the $\xi_2$ prediction of this model (solid curves, with lower bias predicting lower $\xi_2/\xi_{2,{\rm lin}}$) to the same simulation results as in Fig.~\ref{fig:streaming1}.  The fit is excellent for the HOD subsample.  The inaccuracy for the other mass bins can be traced to the insufficient accuracy of the perturbation theory prediction for $v_{12}(r)$ and its slope on small scales.}
\end{figure}

By combining the results in Section \ref{sec:PT}, using Eq.~\ref{streamingeqn}, we have an analytic prediction for $\xi_{0,2,4}$.  Because $\xi_2$ contains almost all of the available information on $f\sigma_8$ (see Fig.~\ref{fig:fisher1}), we focus on that quantity here.  Fig.~\ref{fig:PTstreamxi2} shows the same simulation results as in Fig.~\ref{fig:streaming1} for $\xi_2$ (symbols with errors) while the lines show the prediction of the perturbation theory scale-dependent Gaussian streaming model.  While the agreement is excellent for the HOD halo subsample, the accuracy of the model is worse than 2 per cent for scales smaller than $\sim 40\,h^{-1}$Mpc.
This disagreement can be traced almost entirely to insufficient accuracy in the perturbation theory prediction
for $v_{12}(r)$.  Replacing the perturbation theory prediction for $v_{12}(r)$ with the simulation results, keeping all
other terms fixed, dramatically improves the agreement with simulations.  This explains why $\xi_2$ for the HOD
model agrees so well for the HOD subsample -- it is for this sample that the perturbation theory model for $v_{12}$
agrees best with the simulations.
Eq.~\ref{vexpand} demonstrates that even in the linear limit, both $v_{12}(r)$ and its derivative contribute to $\xi_2$, so the degradation in accuracy from $v_{12}(r)$ to $\xi_2$ is primarily because the slope of $v^{PT}_{12}(r)$ is inaccurate for the low and high mass halo sample, but coincidentally correct for the HOD subsample.  We plan to explore other perturbation theory approaches for $v_{12}(r)$ in a future paper in hopes of improving the accuracy of the model.

\section{Redshift Dependence}
\label{zdep}

\begin{figure}
\centering
\resizebox{3.6in}{!}
{\includegraphics{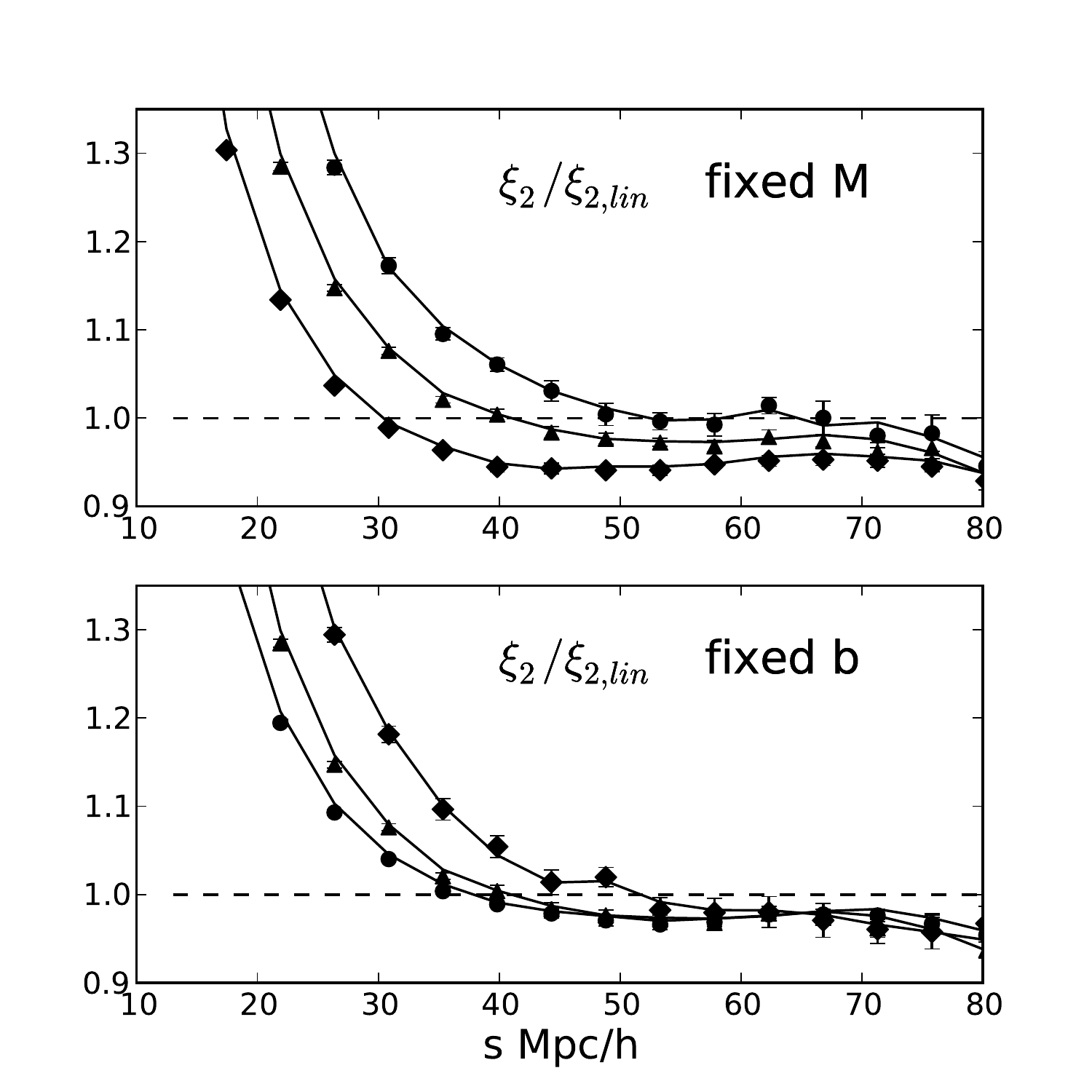}}
\caption{\label{fig:redshiftdep}  To check the redshift dependence of our model, we show $\xi_2/\xi_{2,lin}$ at different simulation snapshots $z=1$ (circles), $z=0.55$ (triangles), and $z=0$ (diamonds).  In the upper panel, we select all halos above a fixed minimum halo mass $M_{min} = 10^{13.387} \; h^{-1} M_{\sun}$.  In the lower panel, we vary the minimum halo mass included in the sample such that the sample bias remains fixed at 2.8; the $z=0.55$ sample (the `high subsample' in previous plots) is the same in both panels.  The solid lines are the predictions from the Gaussian scale-dependent streaming model with input real space clustering and velocity statistics measured from the $N$-body simulations (as in Fig.~\ref{fig:streaming1}).}
\end{figure}

Though we have focused on $z=0.55$, Fig.~\ref{fig:redshiftdep} shows that the scale-dependent Gaussian streaming model works extremely well for massive halos both at $z=1$ and $z=0$.  In the upper panel, we compare samples containing all halos above $M_{\rm min} = 10^{13.387} \; h^{-1} M_{\sun}$, for which the best fit bias values are $b_{LPT} = 3.83, 2.79$ and 1.88 (in order of decreasing redshift).  In the lower panel, we vary $M_{\rm min} = 10^{12.9}, 10^{13.387}, 10^{13.95} \; h^{-1} M_{\sun}$, so that the bias of the samples is fixed at $b_{LPT}\simeq 2.8$.  The redshift dependence of $\xi_2/\xi_{2,{\rm lin}}$ on small scales is opposite in the two panels, which can be understood simply in terms of the absolute real space clustering amplitude that enters the convolution in Eq.~\ref{streamingeqn}.  At the fixed value of $M$ we have chosen, $\xi^r \propto b(z) \sigma_8(z)$ decreases as structure grows, whereas at fixed bias, $\xi^r$ increases with time as $\sigma_8(z)$.  We use this example to emphasise that the non-linear corrections are not necessarily smaller at higher redshift, since more highly biased objects are often being selected.  We have checked the behaviour of the perturbation theory predictions at $z=0$ and $z=1$ for the real space statistics examined in Section \ref{sec:PT}, and find good agreement with naive expectations.  At fixed $b=2.8$, the sudden upturn in the LPT prediction for $\xi^r$ occurs at increasingly large $r$ as structure evolves (22, 26, and $31\,h^{-1}$Mpc at z=1, 0.5, 0, respectively).  At fixed $z$, the upturn increases with $b$, as can be seen in Fig.~\ref{fig:xirealPTfit}.  We find similar results when comparing pairwise halo velocity statistics; in general, the perturbation theory predictions are better at higher redshift and lower bias.  However, the ``sweet spot'' in the perturbation theory prediction at $b=1.9$ in Fig.~\ref{fig:vstats} persists at $z=0$ as well, though the halo mass range at fixed $b$ depends on $z$.

\section{Bias Dependence}
\label{biasdep}

\begin{figure}
\centering
\resizebox{3.6in}{!}
{\includegraphics{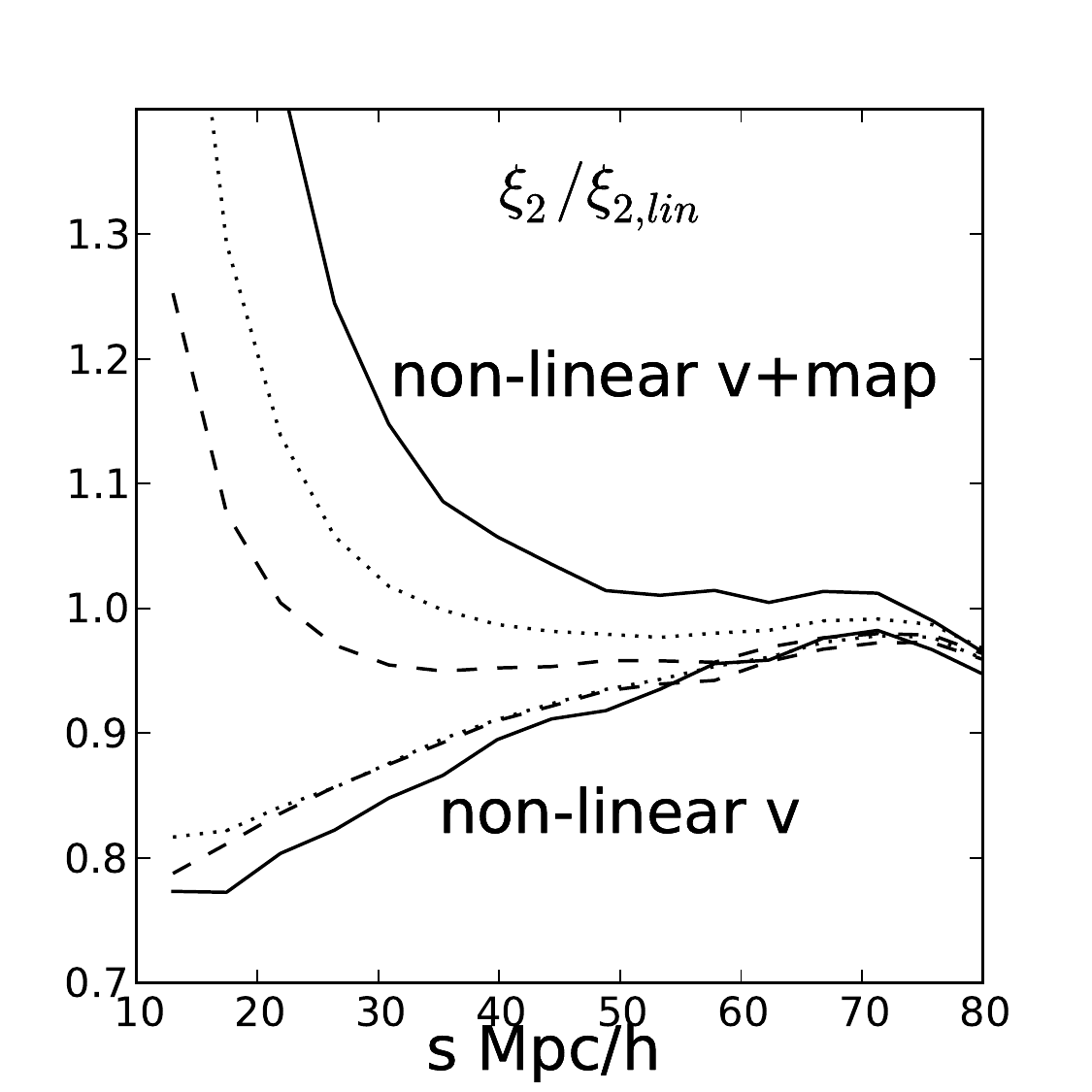}}
\caption{\label{fig:biasdep}  The predictions for $\xi_2$ for halos (b=2.8, solid; b=1.9, dotted; b=1.4, dashed) for two different mappings between real and redshift space using the simulation results for the non-linear real space clustering and velocity statistics.  The first mapping (lower curves at $s=30\,h^{-1}$Mpc) is equivalent the mapping assumed in the Kaiser formula (our Equations \ref{fisherlinearxi} through \ref{vdispexpand}), while the second includes the dominant correction term (Eq.~\ref{b3term}) from the non-linear mapping, which scales like $b^3$.}
\end{figure}

As we have emphasised, there are two relatively large corrections to the Kaiser prediction for $\xi_2$.  In this section we illustrate explicitly how they depend on bias.  The suppression of the halo velocity statistics relative to linear theory lowers the amplitude of $\xi_2$, while the non-linear mapping between real and redshift space increases the amplitude of $\xi_2$ in the bias range we have studied (see Fig.~\ref{fig:streaminglin}).  To illustrate the effect of the former, we use the Kaiser limit mapping between real space clustering and velocity statistics (Equations \ref{fisherlinearxi} through \ref{vdispexpand}), but input the real space $N$-body simulation results for these quantities rather than the linear theory expectations.  The results for the halo subsamples in Table \ref{table:halos} are the lower three curves in Fig.~\ref{fig:biasdep}.  The weak dependence of the non-linear corrections on the velocity statistics (Fig.~\ref{fig:vstats}) translates into a relatively weak dependence on halo bias on $\xi_2$.  In contrast, the effect of the non-linear mapping depends strongly on halo bias.  To see this explicitly, one can expand Eq.~\ref{streamingeqn} by assuming that the pairwise velocity PDF $\mathcal{P}(v_z ; {\bf r})$ is a smooth and slowly varying function of ${\bf r}$; this procedure will be more accurate at small $\mu$, where a smaller range of real space separations contribute pairs at a given redshift space separation.  Eq.~53 of \citet{Sco04} does this in the case of the exact Gaussian result, and the same terms (along with many others) appear when the expansion is performed on our Eq.~\ref{streamingeqn}.  We have verified that the dominant non-linear correction term for $\xi_{0,2}$ in our bias range comes from the term $-d/dy [\xi v_{12}]$:
\begin{equation} \label{b3term}
\Delta \xi_{mapping}(r) = (\mu^2-1) \frac{\xi(r) v_{12}(r)}{r} - \mu^2\frac{d[\xi(r) v_{12}(r)]}{dr}.
\end{equation}
The upper curves in Fig.~\ref{fig:biasdep} are the same as the lower ones, but with this extra term included to approximate the non-linear mapping step; these predictions are in reasonable agreement with Eq.~\ref{streamingeqn}, but performing the full integral is a noticeably better fit to the simulation results.  What we wish to emphasise is that the non-linear mapping produces a term (Eq.~\ref{b3term}) that contributes to $\xi_0$ and $\xi_2$ and scales like $b^3$.  This is in disagreement with the recent results of \cite{tang/kayo/takada:2011}, who use a non-linear correction term equivalent to the $\left < \delta_i \delta_j {\bf v}_{4-i-j}\right >$ contribution in our Eq.~\ref{vinfallPT}.  We can see why that term (the dotted curve in Figure \ref{fig:PTtot}) provides a reasonable fit to their simulation results at one value of $b$: its shape roughly mimics our non-linear mapping term that dominates on small scales.  However, our more detailed analysis demonstrates that many other terms are of comparable size to the one considered in \citet{tang/kayo/takada:2011}.

\citet{okumura/jing:2011} find that the value of $\beta$ recovered from massive halos $b \gtrsim 1.5$ is relatively close to the expected linear value, but lower mass halos recover a smaller value compared with linear theory.  Fig.~\ref{fig:biasdep} illustrates why: for our central galaxy sample (dotted curves), the non-linear effects of velocity suppression and real-to-redshift space mapping approximately cancel for $s > 30\,h^{-1}$Mpc; at low halo bias, we expect the non-linear mapping corrections to be small, and the measured $\xi_2$ should be closer to the lower curves.  Of course, the bias where this near-cancellation occurs will depend on redshift, and because of the $b^3$ correction term, it will only be true in a limited range of bias values.

To be more quantitative, for the halo bias range we have studied ($b=1.4-2.8$), fitting the Kaiser formula to $\xi_0$ and $\xi_2$ to derive constraints on $b$ and $f$ on scales $r \sim 30\,h^{-1}$Mpc will bias the constraints on $f$ by +2, -6, and -10 per cent for $b_{lin} =$ 2.67, 1.84, 1.41, respectively, under the assumption that the smallest scales included dominate the signal-to-noise.  These biases are already at the level of current statistical errors.

While our simulations cannot reach halos with $b \sim 1$, our analysis can shed some light on what behaviour to expect for $b \approx 0.8 - 1.2$ halos of relevance to the WiggleZ survey \citep{blake/etal:2011} and more closely related to the perturbation theory studies for matter.  Fig.~\ref{fig:streaminglin} shows that the non-linear mapping should be a small correction for $r \gtrsim 30\,h^{-1}$Mpc, but that it still amplifies $\xi_2$ on smaller scales.  If we evaluate our perturbation theory predictions for $b=1$, we find that the total correction to $v_{12}(r)$ can be well-approximated by only the $P_{\delta \theta}$ term; however, the higher order terms still dominate the corrections to the velocity dispersions.  Therefore, while the bispectrum terms should not be negligible, fitting formulae based on Eq.~\ref{scoccPT} with $\sigma_v^2$ treated as a free parameter capture at least some of the relevant non-linear corrections and can absorb the rest into $\sigma_v^2$; even while providing a good fit to the data, we are not guaranteed that the underlying peculiar velocity field amplitude will be recovered at the few per cent level, especially when fitting down to $k_{max} = 0.3 \; h$ Mpc$^{-1}$ (i.e., $s_{min} \approx 10\,h^{-1}$Mpc).

\section{Discussion}
\label{discussion}

In contrast to many recent theoretical investigations of redshift space distortions, which have focused on the matter density field and/or performed analyses in Fourier space, in this paper we focus on the redshift space clustering of dark matter halos in configuration space, and use $67.5\,h^{-3}\,{\rm Gpc}^3$ of $N$-body simulations to make precise measurements of $\xi_{0,2,4}$ as a function of halo bias.  In our modelling we focus on two distinct corrections to the linear theory predictions: the non-linear mapping between real and redshift space, and the non-linearity of the halo pairwise velocities.  We find both corrections to be important on the quasilinear scales of interest to this work ($\sim 30-80\,h^{-1}$Mpc).

To model the non-linear real to redshift space mapping of pairs, we take a non-perturbative approach and employ the scale-dependent Gaussian streaming model, Eq.~\ref{streamingeqn}, in which the pairwise velocity probability distribution function (PDF) is assumed to be Gaussian, but where the pairwise velocities have a mean and dispersion that depend on the pair separation distance $r$ and the angle of the pair separation vector with the LOS.  A similar model has been used with some success on somewhat smaller scales \citep{tinker/weinberg/zheng:2006,tinker:2007} in the context of the halo model.  Fig.~\ref{fig:streaminglin} shows that this model significantly enhances both $\xi_{2,4}$ on quasilinear scales for real space statistics expected in linear theory.  When we know perfectly the real space clustering and velocity statistics, Fig.~\ref{fig:streaming1} shows that this model is accurate at the $\lesssim 2$ per cent level for $s > 25\,h^{-1}$Mpc, i.e., at about the level demanded by ongoing experiments like BOSS.

For the first time (to our knowledge), we have computed the next-to-leading order corrections to pairwise mean infall velocities and dispersions for linearly biased halos as a function of real space separation in standard perturbation theory.  While we find relatively good agreement between our calculations and halo samples drawn from a large volume of N-body simulations (see Fig.~\ref{fig:vstats}), when used as input into the streaming model, there remain offsets at the several percent level for $s \leq 40\,h^{-1}$Mpc.  We are able to trace the discrepancy to the scale-dependent inaccuracy of the perturbation theory prediction for halo infall velocities; note that even in linear theory, the redshift space quadrupole $\xi_2$ depends on the derivative $dv_{12}(r)/dr$.  In future work we hope to explore other perturbation theory schemes as a means to improve the level of accuracy of the mean pairwise velocity perturbation theory prediction; more complex biasing schemes may also improve agreement \citep{desjacques/sheth:2010}.

Both of the corrections we described above have significant higher order contributions that depend on the halo bias.  In Section \ref{biasdep}, we show that while the bias dependence is detectable but weak for the non-linear velocity corrections, the dominant correction for the non-linear mapping from real and redshift space scales as $b^3$.  These effects have opposite signs in the halo bias range we have studied, so that for some limited range of halo bias and scale, they approximately cancel.  Our findings demonstrate, in line with other recent works, that a model of the form in Eq.~\ref{scoccPT} that only includes two-point corrections cannot accurately describe the dependence of redshift space halo clustering on bias.

Finally, we note that in order to infer $f\sigma_8$ from redshift space distortions in halo clustering, one must make the assumption that the bias inferred from real space clustering is the same one that determines the halo pairwise infall velocity amplitude.  Our large volume of $N$-body simulations allows us to confirm this assumption at the per cent level on large scales, once we incorporate the effects of non-linear growth using perturbation theory.

Before our model can be applied to analysing real galaxy surveys, the large velocity dispersions of satellite galaxies must be accounted for separately in the model.  We are cautiously optimistic that these additional corrections, at least for $\xi_2$, will be relatively small.  In a preliminary study, we found that $\xi_2(s=30\,h^{-1}{\rm Mpc})$ is damped by $\lesssim 2$ per cent when satellites from the best fit HOD of \citet{white/etal:2011} are included. However, our model prediction for $\xi_4$ is worse than for the halo-only samples.  We reserve this line of research for a future work. 

\section*{Acknowledgements}
BAR thanks Jeremy Tinker and Ravi Sheth for insightful discussions.  Support for this work was provided by NASA through Hubble Fellowship grant 51280 awarded by the Space Telescope Science Institute, which is operated by the Association of Universities for Research in Astronomy, Inc., for NASA, under contract NAS 5-26555.
MW is supported by the NSF and NASA.
The simulations used in this paper were analysed at the National Energy
Research Scientific Computing Center, the Shared Research Computing Services
Pilot of the University of California and the Laboratory Research Computing
project at Lawrence Berkeley National Laboratory.

\appendix

\section{Perturbation Theory Calculation Details}
\label{ptcalcs}

We refer the reader to \citet{bernardeau/etal:2002} for an introduction to standard perturbation theory.  In this appendix we provide explicit expressions for all terms contributing up to fourth order in the linear density field $\delta_1({\bf k})$ to pairwise mean velocities and dispersions for linearly biased tracers.  Our results are presented using the Fourier convention given in Eq.~\ref{eq:fourier}.

\subsection{Mean halo infall velocities}

Terms in Eq.~\ref{eq:vinfall} that depend on only one $\delta$ can be written in terms of the perturbation theory density-velocity cross power $P_{\delta\theta}$:
\begin{eqnarray}
2b\left(\left\langle\delta_1({\bf x}) {\bf v}_1({\bf x}+{\bf r})\right\rangle + \sum_{i>0} \left\langle\delta_i({\bf x}) {\bf v}_{4-i}({\bf x}+{\bf r})\right\rangle\right) = \nonumber \\
-\hat{r} \frac{fb}{\pi^2} \int dk k P_{\delta \theta}^{PT}(k) j_1(kr) 
\end{eqnarray}
We evaluate $P^{PT}_{\delta \theta}$ using the publicly available {\em Copter} code \citep{carlson/white/padmanabhan:2009}.  Strictly speaking, the perturbation theory result for $P_{\delta \theta}(k)$ becomes negative at large $k$; we simply truncate the integral over $k$ at $k_{max} = 2$ $h$ Mpc$^{-1}$, where linear theory predictions are already well converged.

There are three distinct contributions of the form $\left\langle\delta_i \delta_j {\bf v}_{4-i-j}\right\rangle$:
\begin{eqnarray}
\left\langle\delta_1({\bf x}) \delta_2({\bf x}+{\bf r}) v_1({\bf x}+{\bf r})\right\rangle = \nonumber \\
2 \int \frac{d^3k_2 d^3k_3 if {\bf k_3}}{(2\pi)^6 k_3^2} P(|{\bf k}_2+{\bf k}_3|) P({\bf k}_3) F_2(-{\bf k}_3, {\bf k}_2 + {\bf k}_3) e^{i ({\bf k}_2+{\bf k}_3) \cdot {\bf r}} \label{d1d2v1eqnv1}\\
\left\langle\delta_1({\bf x}) \delta_1({\bf x}+{\bf r}) v_2({\bf x}+{\bf r})\right\rangle = \nonumber \\
2 \int \frac{d^3k_2 d^3k_3 i f {\bf k_3}}{(2\pi)^6 k_3^2} P(|{\bf k}_2+{\bf k}_3|) P({\bf k}_2) G_2(-{\bf k}_2, {\bf k}_2 + {\bf k}_3) e^{i ({\bf k}_2+{\bf k}_3) \cdot {\bf r}}  \label{d1d1v2eqnv1} \\
\left\langle\delta_2({\bf x}) \delta_1({\bf x}+{\bf r}) v_1({\bf x}+{\bf r})\right\rangle = \nonumber \\
2 \int \frac{d^3k_2 d^3k_3 i f {\bf k_3}}{(2\pi)^6 k_3^2} P({\bf k}_2) P({\bf k}_3) F_2({\bf k}_2, {\bf k}_3) e^{i ({\bf k}_2+{\bf k}_3) \cdot {\bf r}}  \label{d2d1v1eqnv1}
\end{eqnarray}
where $F_2$ and $G_2$ are the standard perturbation theory kernels for $\delta_2$ and $\theta_2$, respectively.  

Performing some of the angular integrations, we find
\begin{eqnarray}
\left\langle\delta_1({\bf x}) \delta_2({\bf x}+{\bf r}) v_1({\bf x}+{\bf r})\right\rangle +  \left\langle\delta_1({\bf x}) \delta_1({\bf x}+{\bf r}) v_2({\bf x}+{\bf r})\right\rangle  = \nonumber \\
= -\frac{f \hat{{\bf r}}}{4\pi^4} \int_0^{\infty} \int_0^{\infty} du \; dy \; u^4 P(u) P(yu) j_1(ur) \times \nonumber \\
\frac{2y(-19+24y^2-9y^4)+9(-1+y^2)^3 T}{168y} \label{d1d1v2eqnv2} \\
\left\langle\delta_2({\bf x}) \delta_1({\bf x}+{\bf r}) v_1({\bf x}+{\bf r})\right\rangle = 
-\frac{ f \hat{{\bf r}}}{4\pi^4} \int_0^{\infty} u^4 j_1(ur) du \times \nonumber \\
\int_0^{\infty}P(u y) dy \int_{-1}^1 dw  P(u\sqrt{1+y^2-2yw}) \frac{3yw-10y w^3 + 7w^2}{14(1+y^2-2yw)} \label{d2d1v1eqnv2}
\end{eqnarray}
where we have defined $T={\tanh}^{-1}(2y/[1+y^2])$.
Note the last integrand can also be expressed as a sum of products of one-dimensional integrals that each depend on $r$.  We prefer the expression above, which can be combined with Eq.~\ref{d1d1v2eqnv2} before doing the final integral over $j_1(ur)$.  We caution the reader that the integrals are only well-behaved in combination; for instance, $\left\langle\delta_1({\bf x}) \delta_2({\bf x}+{\bf r}) v_1({\bf x}+{\bf r})\right\rangle$ and $\left\langle\delta_1({\bf x}) \delta_1({\bf x}+{\bf r}) v_2({\bf x}+{\bf r})\right\rangle$ do not converge individually for a CDM-like power spectrum, where $P(k) \sim k^{-2.6}$ at large $k$.  Moreover, the integral over $x=uy$ in Eq.~\ref{d1d1v2eqnv2} yields a constant at large $u$, which should then be integrated against $u^3 P(u) j_1(ur)$; combining with Eq.~\ref{d2d1v1eqnv2} before performing the integral on $u$ yields in integrand that behaves like $u^{-0.9} j_1(ur)$ at large $u$, which converges.

\subsection{Halo velocity dispersions}

In this section we evaluate the contributions to $\sigma_{12}^2(r,\mu^2)$, given in Equations \ref{eq:vdisp} to \ref{v2r6}.  We first introduce some shorthand notation:
\begin{eqnarray}
J(\mu_{\ell r}^2, kr) = \mu_{\ell r}^2 \left(j_0(kr) - \frac{2j_1(kr)}{kr}\right) + (1 - \mu_{\ell r}^2)\frac{j_1(kr)}{kr}\\
I_{n,m}(y) = \int_0^{\pi} sin^n(\theta) cos^m(\theta) e^{iycos(\theta)} d\theta.
\end{eqnarray}
The higher-order contributions to line \ref{v2r1} can be expressed in terms of the perturbation theory result of the velocity divergence power spectrum:
\begin{equation}
\left\langle v^{\ell}(\bf{x}) v^{\ell}({\bf x}+{\bf r})  \right\rangle = \frac{f^2}{2\pi^2} \int dk P_{\theta \theta}^{PT}(k) J(\mu_{\ell r}^2, kr),
\end{equation}
which is equivalent to Equations \ref{defvcorrlin} - \ref{psiparallel} at linear order.  We evaluate $P_{\theta \theta}^{PT}$ with {\em Copter} as well, again truncating the integral over $k$ at $k_{max} = 2\,h$ Mpc$^{-1}$, where linear theory predictions are already well converged.  
The term in line \ref{v2r2} adds a constant to $\sigma_{12}^2(r,\mu_{\ell r}^2)$:
\begin{eqnarray}
\left\langle\delta({\bf x}) (v^{\ell}({\bf x}))^2\right\rangle = 2 \left\langle\delta_1({\bf x}) v^{\ell}_1({\bf x}) v^{\ell}_2({\bf x})\right\rangle + \left\langle\delta_2({\bf x}) (v^{\ell}_1({\bf x}))^2 \right\rangle \nonumber \\
=\frac{f^2}{6\pi^4} \int k^3 dk \; dy \; P(k) P(ky) \times \nonumber \\
\frac{2y(3+8y^2-3y^4)+3(-1+y^2)^3 T}{56y},\label{dxvx2n2}
\end{eqnarray}
where again neither term converged individually.
There are five distinct contributions to the terms in line \ref{v2r4}:
\begin{eqnarray}
\left(\left\langle\delta({\bf x}) (v^{\ell}({\bf x}+{\bf r}))^2 \right\rangle - 2 \left\langle\delta({\bf x}) v^{\ell}({\bf x}) v^{\ell}({\bf x}+{\bf r}) \right\rangle\right) = \nonumber \\
2\left\langle\delta_1({\bf x}) v_1^{\ell}({\bf x} + {\bf r}) v_2^{\ell}({\bf x} + {\bf r}) \right\rangle + \left\langle\delta_2({\bf x}) (v_1^{\ell}({\bf x} + {\bf r}))^2 \right\rangle \nonumber \\
 -2\left\langle\delta_1({\bf x}) v^{\ell}_1({\bf x}) v^{\ell}_2({\bf x} + {\bf r}) \right\rangle  -2 \left\langle\delta_1({\bf x}) v^{\ell}_2({\bf x}) v^{\ell}_1({\bf x} + {\bf r}) \right\rangle \nonumber \\ 
 -2 \left\langle\delta_2({\bf x}) v^{\ell}_1({\bf x}) v^{\ell}_1({\bf x} + {\bf r}) \right\rangle.
\end{eqnarray}
Performing as many angular integrals as possible, we are left with the following expressions, which we evaluate numerically.
\begin{eqnarray}
2 \left\langle\delta_1({\bf x}) v^{\ell}_1({\bf x}+{\bf r}) v^{\ell}_2({\bf x}+{\bf r})\right\rangle = \frac{f^2}{2\pi^4} \int P(w) P(wy) w^3 j_0(wr) dw dy \times \nonumber \\
\frac{-18y+66y^3+66y^5-18y^7 + 9(-1+y^2)^4 T}{672y^3} \nonumber \\
-\frac{f^2}{2\pi^4} \mathcal{H}^2 f^2 \int P(w) P(wy) w^3 J(\mu_{\ell, r}^2, wr) dw dy \times \nonumber \\
\frac{2y(9 - 109 y^2 + 63y^4 - 27y^6) + 9(-1+y^2)^3(1+3y^2) T}{672y^3} \\
\left\langle\delta_2({\bf x}) (v^{\ell}_1({\bf x}+{\bf r}))^2 \right\rangle = -\frac{f^2}{32\pi^4} \int k_2 k_3 dk_2 dk_3 P(k_2) P(k_3) \times \nonumber \\
(\frac{20\mu_{\ell r}^2}{7} I_{1,1}(k_2r) I_{1,1}(k_3r) + \nonumber \\
\frac{1}{2}\left(\frac{k_2}{k_3} + \frac{k_3}{k_2}\right)\left(4\mu_{\ell r}^2 I_{1,2}(k_2r) I_{1,2}(k_3r) + (1-\mu_{\ell r}^2) I_{3,0}(k_2r) I_{3,0}(k_3r)\right) \nonumber \\
+\frac{4}{7}(2\mu_{\ell r}^2 I_{1,3}(k_2r) I_{1,3}(k_3r) + I_{3,1}(k_2r) I_{3,1}(k_3r)) \\
-2\left\langle\delta_1({\bf x}) v_1^{\ell}({\bf x}) v_2^{\ell}({\bf x}+{\bf r}) \right\rangle = \frac{f^2}{28\pi^4}\int k^3 dk dy dx J(\mu_{\ell, r}^2, kr) \times \nonumber \\
P(k\sqrt{1+y^2-2yx}) P(ky) \frac{yx + 6x^3y - 7x^2}{1+y^2-2yx} \\
-2\left\langle\delta_1({\bf x}) v_2^{\ell}({\bf x}) v_1^{\ell}({\bf x}+{\bf r}) \right\rangle -2\left\langle\delta_2({\bf x}) v_1^{\ell}({\bf x}) v_1^{\ell}({\bf x}+{\bf r}) \right\rangle= \nonumber \\
-\frac{f^2}{2\pi^4} \int dy k^3 dk J(\mu_{\ell,r}^2, kr) P(k) P(ky) \times \nonumber \\
\frac{-2y(19 - 24y^2 + 9y^4) + 9(-1+y^2)^3 T}{168y} \label{d1v2v1xprn5}
\end{eqnarray}

\label{lastpage}
\end{document}